\documentclass[a4paper]{article}
\usepackage{graphicx}
\usepackage[parfill]{parskip}
\usepackage{amsmath, amssymb, amsfonts}
\usepackage[colorlinks = true,
			citecolor  = blue,
            linkcolor = black]{hyperref}
\usepackage[font={scriptsize}]{caption}
\usepackage[labelfont=bf]{caption} 
\usepackage{algorithm}
\usepackage{algpseudocode}

\usepackage[natbibapa]{apacite}
\usepackage{float}
\usepackage[
top    = 1.0in,
bottom = 1.2in,
left   = 1.0in,
right  = 1.0in]{geometry}
\title{Comparing MCMC algorithms in Stochastic Volatility Models using Simulation Based Calibration}
\author{Benjamin Wee}
\date{\small October 2023\\[0.5cm]{\small Supervised by Professor Catherine Forbes and Dr Lauren Kennedy}}

\begin{document}

\maketitle 

\begin{abstract}
    Simulation Based Calibration (SBC) \citep{talts2020validating} is applied to analyse two commonly used, competing Markov chain Monte Carlo algorithms for estimating the posterior distribution of a stochastic volatility model. In particular, the bespoke ‘off-set mixture approximation’ algorithm proposed by \citet{kim1998stochastic} is explored together with a Hamiltonian Monte Carlo algorithm implemented through Stan \citep{stan}. The SBC analysis involves a simulation study to assess whether each sampling algorithm has the capacity to produce valid inference for the correctly specified model, while also characterising statistical efficiency through the effective sample size. Results show that Stan's No-U-Turn sampler, an implementation of Hamiltonian Monte Carlo, produces a well-calibrated posterior estimate while the celebrated off-set mixture approach is less efficient and poorly calibrated, though model parameterisation also plays a role. Limitations and restrictions of generality are discussed.
    
\end{abstract}

\newpage

\section*{Acknowledgements}
First and foremost I would like to thank my supervisors Professor Catherine Forbes and Dr Lauren Kennedy. Your advice, guidance and patience has been invaluable. I am eternally grateful for the opportunity to learn from both of you.

I would also like to acknowledge the Masters research coordinator Professor Param Silvapulle for always checking in and keeping me and my fellow students on track.

Special thanks to Swen for her efforts in proofreading, providing feedback and being a reliable friend who I could talk with about my research.

And to my wife Dana who has supported and encouraged me from the very beginning. Thank you always for your love and patience every single day.

\newpage

\tableofcontents{\protect\newpage}

\section{Introduction}
    Stochastic volatility (SV) models are used in financial econometrics to model and predict the behaviour of financial asset returns. Typically expressed as a non-Gaussian state space model where the conditional variance of the observed return is treated as a latent random variable \citep{hull1987pricing, chesney1989pricing}, such models are difficult to estimate since the likelihood is unavailable in closed form. With more parameters than observations\footnote{We refer to both static unknowns and time-varying latent state variables as 'parameters'.}, SV models are commonly analysed using Bayesian methods obtained via a Markov Chain Monte Carlo (MCMC) algorithm designed to sample from the high dimensional joint posterior distribution. The focus of this paper is to investigate the performance of two competing MCMC algorithms, each applied to a SV model, in terms of the validity and statistical efficiency of the their resulting inference. Performance of these algorithms will also be compared with different parameterisations of the SV model to determine the sensitivity of MCMC calibration to model specification.

    \citet{kim1998stochastic} propose a sampling strategy for the so-called `vanilla' stochastic volatility through an off-set mixture approximation. Conditional on the mixture allocations and static parameters, the forward filter backward sampling algorithm \citep{carter1994gibbs, fruhwirth1995bayesian} is used to jointly sample the latent states while conditionally  conjugate prior distributions are also suggested.  Since then, there have been many advances in statistical computing and algorithm design. In particular, Hamiltonian Monte Carlo (HMC) has been proposed as a generic and efficient sampling method for MCMC. Importantly, the adoption of such new techniques have been made widely available through the development of various open source libraries such as the Stan programming language \citep{stan} and the PyMC library \citep{pymc2023}. As the development of new algorithms rapidly increase, so does the need to develop new methods to assess their output. Developments in statistical workflow are required to test new algorithms as well as compare computational strategies used to estimate increasingly complex models.

    This research analyses and compares the calibration and efficiency of MCMC used to estimate stochastic volatility models using a simulation study. By calibration we refer to a MCMC algorithm that returns correct the posterior, on average, conditional on the correct model specification. The first algorithm is the No-U-turn sampler (NUTS) variant of Hamiltonian Monte Carlo (HMC) as implemented in the Stan programming language \citep{hoffman2014no, betancourt2017conceptual, stan}. The second algorithm replicates Kim Shephard and Chib's (KSC) Gaussian off-set mixture model which is sampled using the Kalman Filter and Metropolis Hastings algorithm. The objective is to analyse the efficiency of the MCMC and their ability to return correct posterior estimates on average.

    Comparisons of these algorithms are conducted using a methodology known as Simulation Based Calibration (SBC) \citep{talts2020validating}. Parameters are drawn from the prior distribution and used to create data sets from the generative stochastic volatility model. Posterior samples are then obtained from each of the MCMC methods conditional on the created data set. Repeating this process multiple times gives insight to how well the MCMC algorithm can estimate the true parameters given the data generating process. The key diagnostic metrics are the effective sample size, to measure the efficiency of the algorithm, and uniform rank statistics to determine calibration of the posterior estimates. 

    One particular challenge to comparing models is the parameterisation. Different parameterisations of a model, whilst mathematically the same, can be very different in terms of sampling difficulty \citep{neal2003slice}. In this work, two different parameterisations of the stochastic volatility model are compared for each sampling method. An algorithm may be sensitive to different parameterisations of the same SV model which may impact MCMC performance as described in \citet{strickland2008parameterisation}.
    
    The key findings of this research are HMC produces more calibrated and efficient posterior estimates relative to KSC's MCMC algorithm on the approximate off-set mixture model. Additionally, MCMC algorithms may perform better or worse based on calibration and efficiency depending on the parameterisation of the model. HMC applied on a reparameterised SV model produces the most calibrated and efficient posterior estimates across all SBC experiments.

    This paper is structured as follows. Section 2 provides the context around this research through the introduction of the stochastic volatility model and a discussion of the limitations of using ad-hoc simulations and MCMC convergence diagnostics to determine posterior calibration (or validity). Section 3 describes the SBC methodology, along with the simulation design and diagnostic metrics used for the SV model. Section 4 details key aspects of the two MCMC sampling approaches under consideration, while Section 5 provides the simulation results. Finally, Section 6 discusses findings, limitations and suggests areas for further research.

\section{Research Context}
This section describes the stochastic volatility model and discusses two examples to motivate the research objectives. Section 2.1 outlines the model specification and priors, Section 2.2 explores challenges with inference and convergence diagnostics when the true model is unknown (no ground truth) and Section 2.3 discusses the limitations and pitfalls of running only a single simulation. Section 2.4 lays out the research objective given these motivating examples.

\subsection{Stochastic Volatility}
    The model of interest is the discrete time, univariate stochastic volatility model estimated by \citet{kim1998stochastic} using a Bayesian framework. SV is expressed as a state space model which links the observed (logarithmic) return of a given asset over period $t$, denoted $y_t$, to its latent conditional variance, $h_{t+1}$. This model is given by the following expressions
    \begin{align}
    y_t =& \space \,\exp(h_t/2) \varepsilon_t \\
    h_{t+1} =& \space \, \mu +\phi(h_t - \mu) + \sigma_{\eta} \eta_t  \\
    h_1 \sim& \space \,\mathrm{N}\left(\mu, \frac{\sigma_{\eta}^2}{1-\phi^2}\right) \\
    \varepsilon_t \sim& \space \,\mathrm{N}(0,1) \\
    \eta_t \sim& \space \,\mathrm{N}(0,1).
    \end{align}
    The measurement equation (1) describes the conditional distribution of $y_t$ given $h_t$, where $h_t$ is the conditional log variance of $y_t$. $h_1$ is a draw from a stationary distribution and the state equation $h_{t+1}$ follows a stationary process governed by the auto-regressive parameter $\phi$ such that $|\phi|<1$. This auto-regressive parameter represents the persistence or stickiness of the log variance and the dispersion parameter $\sigma_{\eta}$ is the constant variance of the states. $\epsilon_t$ and $\eta_t$ are standard normal white noise shocks and are uncorrelated with each other.

    KSC assign certain independent prior distributions and hyperparameters to the static parameters, with conjugate priors for $\mu$ and $\sigma^2$, given by
    \begin{align}
    \mu \sim& \space \,\mathrm{N}(0, 10) \\
    \sigma_{\eta}^2 \sim& \space \, \mathrm{IG}(shape=5/2, scale=(0.01\times 5) / 2) \\
    \phi^{\ast} \sim& \space \,\mathrm{Beta}(20, 1.5) \\
    \phi =& \space \, 2\phi^{\ast} - 1.
    \end{align}
    The prior distribution for $\phi$ is referred to as a stretched beta distribution, as it corresponds to a transformed beta random variable, $\phi^*$, with support (-1, 1).

\subsection{Challenge 1: Comparing posteriors with no ground truth}
    The stochastic volatility model can be estimated using a variety of sampling strategies and MCMC algorithms. Convergence metrics assessed from the same MCMC output to be used for posterior inference are commonly used to check the performance of the MCMC chains. Most often, the effective sample size for each parameter is produced, as it indicates the number of approximately independent draws obtained from the dependent Markov chain for a target parameter \citep{gelman2013bayesian}. $\hat{R}$, the potential scale reduction factor or measure of variability between different chains \citep{gelman1992inference}, is also computed to gauge whether the MCMC algorithm has converged to the target posterior distribution.

    Convergence diagnostics are useful for identifying when Bayesian computation fail on real data. However, confounding issues may arise when attempting to diagnose the cause of computational problems. Real data is generated from an unknown data generating process. That is, the true parameter and model are unobservable, and it is very likely that the model being fit is not the true model. Therefore, failed diagnostic checks could be the result of attempting to fit a misspecified model, issues with the MCMC algorithm, or both. Furthermore, different sampling strategies will result in different posterior estimates for the same model. Assuming no issues with the computation or the model specification, convergence diagnostics do not provide any guidance to the analyst regarding which posterior estimate is to be preferred.

\begin{figure}[H]
        \centering
        \includegraphics[scale=0.1]{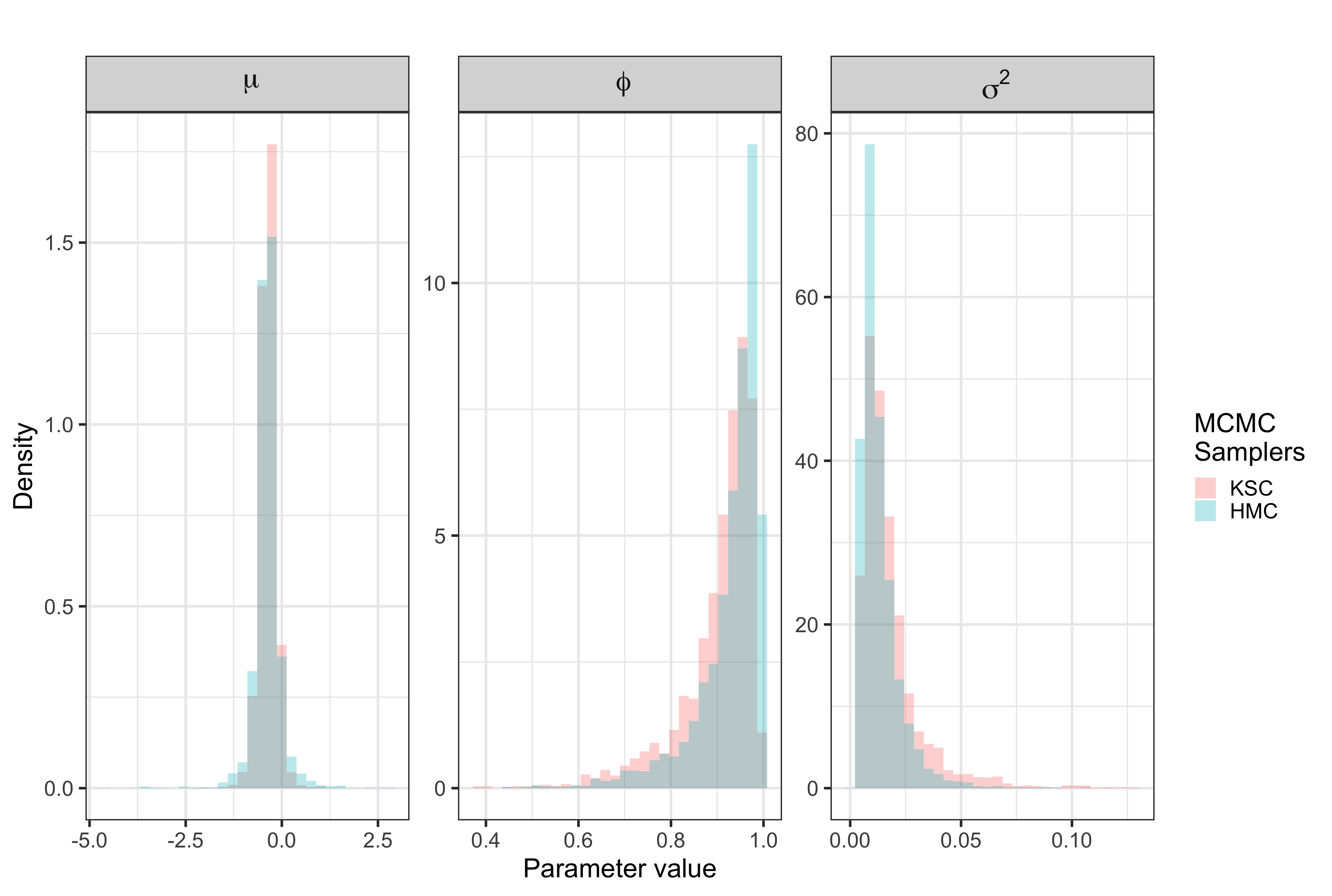}
        \caption{\textbf{Posterior distribution for static parameters fit on the S\&P 500 Index.} Posterior samples are drawn from the HMC (blue) and KSC (red) MCMC algorithms. Marginal posterior densities are presented for $\mu$, $\phi$ and $\sigma^2$. Note that each panel has their own scale for the x and y axis.}
        \label{fig:realdataex}
    \end{figure}

    To illustrate the concerning issues, the posterior distribution associated with the stochastic volatility model in (1)-(5) and a set of real financial returns is estimated using two different MCMC algorithms. The data are demeaned daily (logarithmic) returns on the S\&P 500 Index over the period from 1 January 2023 - 10 September 2023 with sample size $T=171$. Figure \ref{fig:realdataex} displays the marginal posterior estimates for the static parameters $\mu$, $\phi$ and $\sigma^2$, with corresponding summary statistics in Table \ref{tab:realdata}. Overall the two methods are similar in skew and shape. However, closer inspections suggests that $\phi$ has a fatter left tail for the KSC sampler and $\sigma^2$ is also heavier in the right tails relative to HMC. This is evident in the quantiles with the 25th quantile for $\phi$ being smaller at 0.875 for KSC compared to 0.910 in HMC. Furthermore, HMC has taller peaks at their mode relative to KSC with a tighter spread. The uncertainty intervals created from each posterior sample would also be quite different and lead to different results. However, we can only conclude that there are differences between these methods, not which is correct.

    \begin{table}[H]
        \centering
        \begin{tabular}{|c|c|c|c|c|c|c|c|} \hline 
        Parameter&  MCMC&Min& q25&  Median& Mean & q75&Max\\ \hline 
        $\mu$&  KSC&-2.03 & -0.472 & -0.354 & -0.355 & -0.231 & 1.31 \\
     $\mu$&  HMC&-4.55 & -0.508 & -0.373 & -0.370 & -0.229 &2.86  \\\hline 
     $\phi$&  KSC&0.384 & 0.875 & 0.928 & 0.902 & 0.958 & 0.997 \\
     $\phi$&  HMC&0.438 & 0.910 & 0.953 & 0.929 & 0.977 &1.00  \\ \hline 
     $\sigma^2$&  KSC&0.00237 & 0.00891 & 0.0138 & 0.0178 & 0.0210 & 0.130 \\ 
     $\sigma^2$&  HMC&0.00209 & 0.00731 & 0.0105 & 0.0131 & 0.0159 &0.107 \\ \hline
        \end{tabular}
        \caption{\textbf{Summary statistics for HMC and KSC algorithms}. Both sets of estimates appear reasonable but it is unclear from estimates on real data which estimate is closer to the truth.}
        \label{tab:realdata}
    \end{table}
    
\subsection{Challenge 2: Limitations of a single simulation}
    Diagnosing problems with the approximate posterior from models fit to real data is difficult because we do not observe the true model. A strategy around this is to evaluate a model and algorithm on simulated data. One approach is to simulate data from a generative model using known parameters. Then fit the same model on the simulated data and see if the true parameters can be recovered. This gives us the benefit of defining the true parameters of the data generating process to be estimated. If the model and algorithm cannot adequately capture the true parameter, then we cannot be confident that it will provide reliable estimates on real data.

    Simulation is particularly important in the SV model as it has latent parameters. \citet{gelman2020bayesian} discuss that simulation is the only way we can check inference on latent variables. This is critical for the SV model since the underlying framework is a state space model with latent log variance parameters. Latent variables are unobserved in real data and are only estimated in the context of the model. Simulation gives control over the data generating process which reveals what the model can infer about the latent variables. 

    Simulations enable the analyst to check whether a model and corresponding simulation algorithms appropriately estimate the `true' data generating process. However, there are limitations to what can be learned from a single simulation. Even for a correctly specified model, there is always a small probability that the true parameter is in the tails of the corresponding (exact) marginal posterior distribution. \cite{talts2020validating} make the point that a single simulation does not provide sufficient information about the inference made by an algorithm. As discussed in their paper, a single simulation may conclude ``that an incorrectly coded analysis worked as desired, while a correctly coded analysis failed''. 

    Figure 2 shows marginal posterior distributions given a simulated data set of 1000 returns generated by the SV model. The true parameters are $\mu=-0.389$, $\sigma^2_{\eta}=0.0115$ and $\phi=0.547$, and the model is fit using Stan's implementation of HMC, the No-U-Turn sampler, with 1000 post burn-in draws from the joint posterior distribution. The 95\% credible intervals for $\mu$ and $\sigma^2_{\eta}$ cover the true parameter. The true parameter for $\phi$ however, is in the tails and outside the interval. Such an analysis may incorrectly conclude that the model fails to adequately estimate the $\phi$ parameter. However, it may be the case that the posterior distribution is correctly calculated using this algorithm and the results may be due to the features of this specific simulated data set.

    \begin{figure}[h]
        \centering
        \includegraphics[scale=0.1]{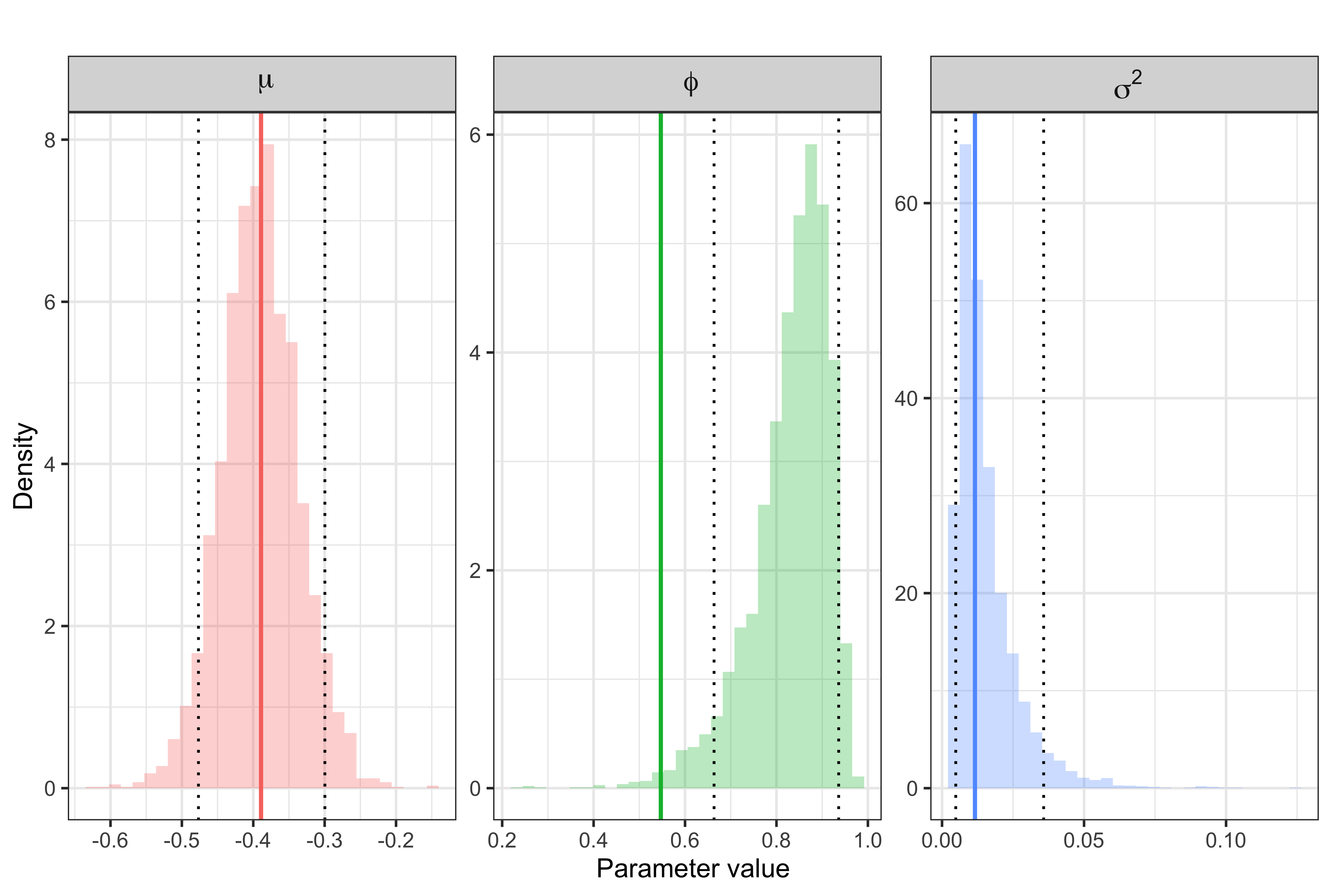}
        \caption{\textbf{HMC posterior samples from SV model fit on known data generating process}. Marginal posterior densities are presented for $\mu$, $\phi$ and $\sigma^2$. Vertical solid line represents true parameter and dotted lines are 95\% credible intervals around the mean. Note that each panel has their own scale for the x and y axis. The parameter $\phi$ falls outside the credible interval whereas $\mu$ and $\sigma^2$ stay inside.}
    \end{figure}

\subsection{Research Goal}
    The objective of this research is to evaluate the calibration of the HMC and KSC algorithms used to estimate SV models through a simulation study\footnote{The software used to produce all results and plots in this research are the R v4.2.3\citep{rlang}, Python v3.8.13 \citep{10.5555/1593511} and Stan v2.31.0 \citep{stan} programming languages, including open source R packages (tidyverse v1.3.2 \citep{tidyverse}, quantmod v0.4.20 \citep{quantmod}, cmdstanr 0.5.3 \citep{cmdstanr}, posterior v1.4.1 \citep{posteriorr}, bayesplot v1.4.10 \citep{bayesplot}, jsonlite v1.8.4 \citep{jsonlite}, arrow v12.0.1.1 \citep{arrow} and future.apply v1.11.0 \citep{RJ-2021-048}) and python packages (numpy v1.24.4 \citep{harris2020array}, scipy v.1.10.1 \citep{2020SciPy-NMeth}, statsmodels v.0.14.0 \citep{seabold2010statsmodels}, pandas v.2.0.3 \citep{mckinney-proc-scipy-2010} and pyarrow v12.0.1 \citep{arrow})}. As discussed, there are limitations to evaluating MCMC algorithms based on fits to real data and single simulations. To check the calibration of an algorithm, repeated independent simulations are required. This methodology is discussed in the next section. 

    Two parameterisations of the SV model for each sampler will be implemented using this simulation design. Results from the study will also be used to compare both sampling strategies to determine which MCMC approach is most suitable for estimating this model. 

\section{Methodology}
This section introduces the simulation design used to evaluate whether a MCMC algorithm is returning the correct posterior estimates on average (that is, whether a MCMC algorithm calibrated). Section 3.1 discusses the procedure for running Simulation Based Calibration and Section 3.2 defines the diagnostics and metrics for evaluating SBC experiment results.

    \subsection{Simulation Design}
        Simulation Based Calibration (SBC) checks the calibration of posterior estimates generated by MCMC algorithms. SBC is conducted by comparing the distribution of rank statistics to the uniform distribution which arises when an algorithm is correctly calibrated. For each $k$ of $K$ SBC iterations, one value for each parameter is drawn from the prior. Conditional on the drawn parameter values, a data set is generated from the model. Then a sampling method (originally using Stan in \citet{talts2020validating}) is used to obtain posterior samples given this data and the prior. The estimated posteriors are then compared to the true values by calculating the rank statistics. Under a calibrated algorithm, the $K$ rank statistics for each parameter should be uniformly distributed. 

        To illustrate the SBC procedure, let $\theta$ be an arbitrary scalar parameter from a model and $y$ represent a data set. Start with a single parameter $\theta^{sim}$ drawn from the prior distribution
        \begin{align}
        \theta^{sim} \sim \pi(\theta),
        \end{align}
        then generate a data set conditional on $\theta^{sim}$ and the model of interest\footnote{Since all computation presumes a correctly specified model, notation to condition on the model is suppressed.},
        \begin{align}
        y^{sim} \sim \pi (y|\theta^{sim}).
        \end{align}
        Take $B$ draws from the estimated posterior distribution, conditional on this data set, and generated by the MCMC algorithm under investigation (here, HMC or KSC)
        \begin{align}
        \{\theta_1,\dots , \theta_{B}\} \sim \pi (\theta | y^{sim}).
        \end{align}

        The last step is to calculate the rank statistic. The rank statistic for any one-dimensional function of parameters $f:\Theta\rightarrow\ \mathbb{R}$ is defined as
        \begin{align}
        r = \mathrm{rank}(\{f(\theta_1),\dots , f(\theta_{B})\}, f(\theta^{sim})) = \sum_{b=1}^{B}1[f(\theta_{b}) < f(\theta^{sim})].
        \end{align}

        Equations (10)-(13) describes one iteration of SBC. To complete the algorithm, multiple iterations are run and the rank statistics are calculated for each parameter. 
        
        \citet{talts2020validating} prove that the sample of rank statistics for a given parameter follow a uniform distribution over the integers $\{0,1\dots,B\}$ if the MCMC algorithm is calibrated. Therefore, the rank statistics calculated from the SBC procedure are compared to the uniform distribution to determine if the algorithm is returning the correct posterior estimates on average.

        Another key result for calibration is that the posterior averaged over the data and true parameters is equal to the prior distribution. This is shown in the following expression
        \begin{align}
        \pi(\theta) &= \int \int \pi(\theta|y^{sim}) \pi(y^{sim}|\theta^{sim}) \pi(\theta^{sim})d\theta^{sim} dy^{sim} \\
        &= \int \int \pi(\theta|y^{sim}) \pi(y^{sim},\theta^{sim}) d\theta^{sim} dy^{sim}.
        \end{align}
        That is, the average posterior for $\theta$ over the SBC iterations should approximately equal the prior distribution, if the algorithm is calibrated. Therefore, any deviation between the prior distribution and the average posterior draws over the SBC iterations may suggest that the sampling methodology is not producing the correct posterior on average.

    \subsection{Evaluation Metrics}
        In this research the following metrics are chosen:

        \begin{enumerate}
            \item Rank statistics to measure calibration for a given parameter and algorithm.
            
            \item Chi-squared statistics calculated over the $K$ rank statistics to summarise the shape of the distribution and deviation from uniformity.

            \item Effective sample size (ESS) to measure MCMC efficiency.
        \end{enumerate}

        \subsubsection{Rank statistics}
            Rank statistics are used to evaluate the calibration of the MCMC algorithm. If a posterior is well calibrated, then it is expected that over $K$ SBC iterations for each univariate parameter in the model:

            \begin{enumerate}
                \item The distribution of rank statistics is uniform.
                \item The posterior draws on average are equal to the prior distribution.
            \end{enumerate}

\begin{figure}[H]
                \centering
                \includegraphics[scale=0.07]{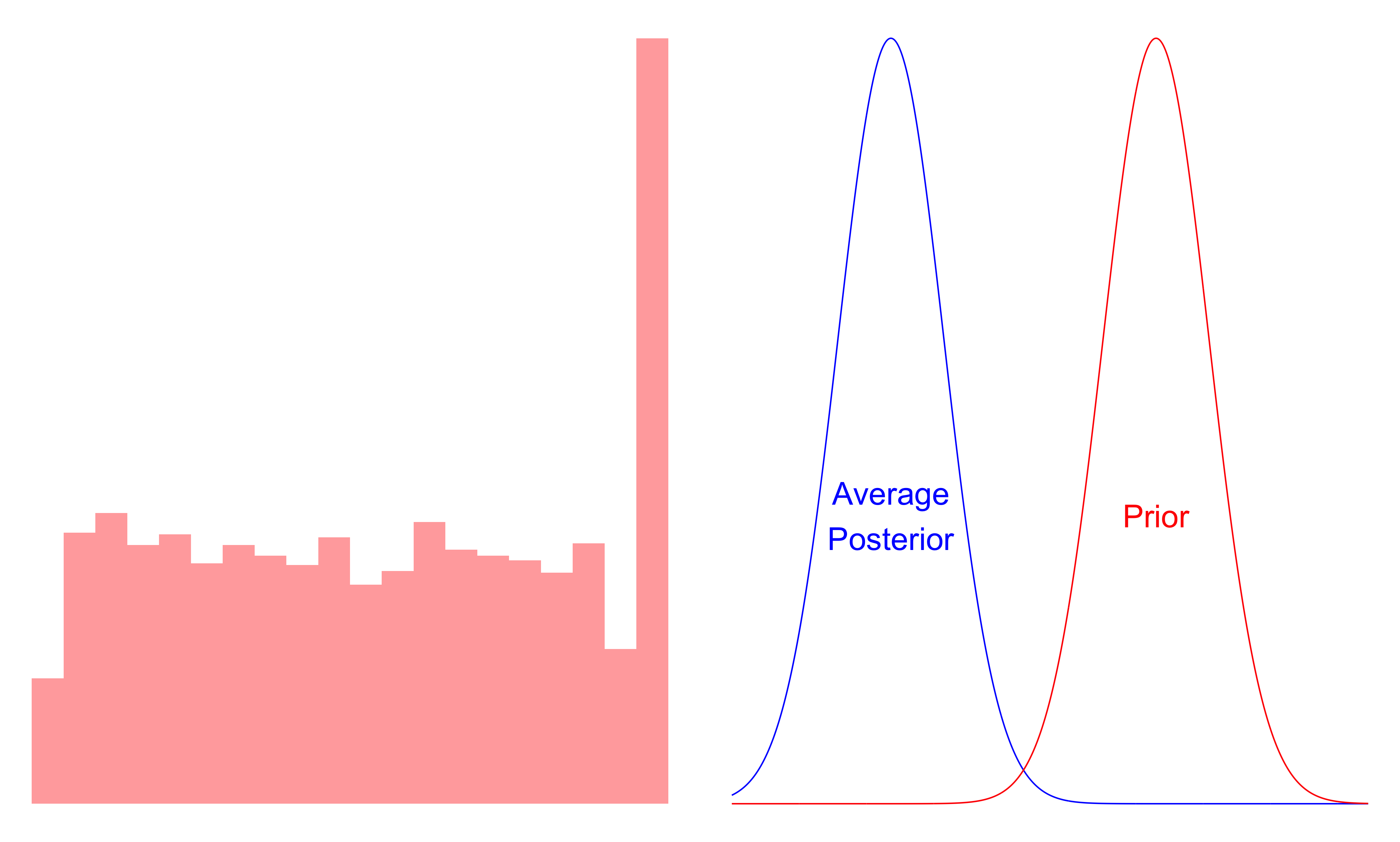}
                \caption{\textbf{Left panel:} Non uniform rank statistics with inflated frequency on the right side. \textbf{Right panel:} The average posterior distribution underestimates the prior distribution. The miscalibration from the rank statistic distribution comes from posterior samples underestimating the prior distribution on average.}
                \label{fig:underestimation}
            \end{figure}

\begin{figure}[H]
                \centering
                \includegraphics[scale=0.07]{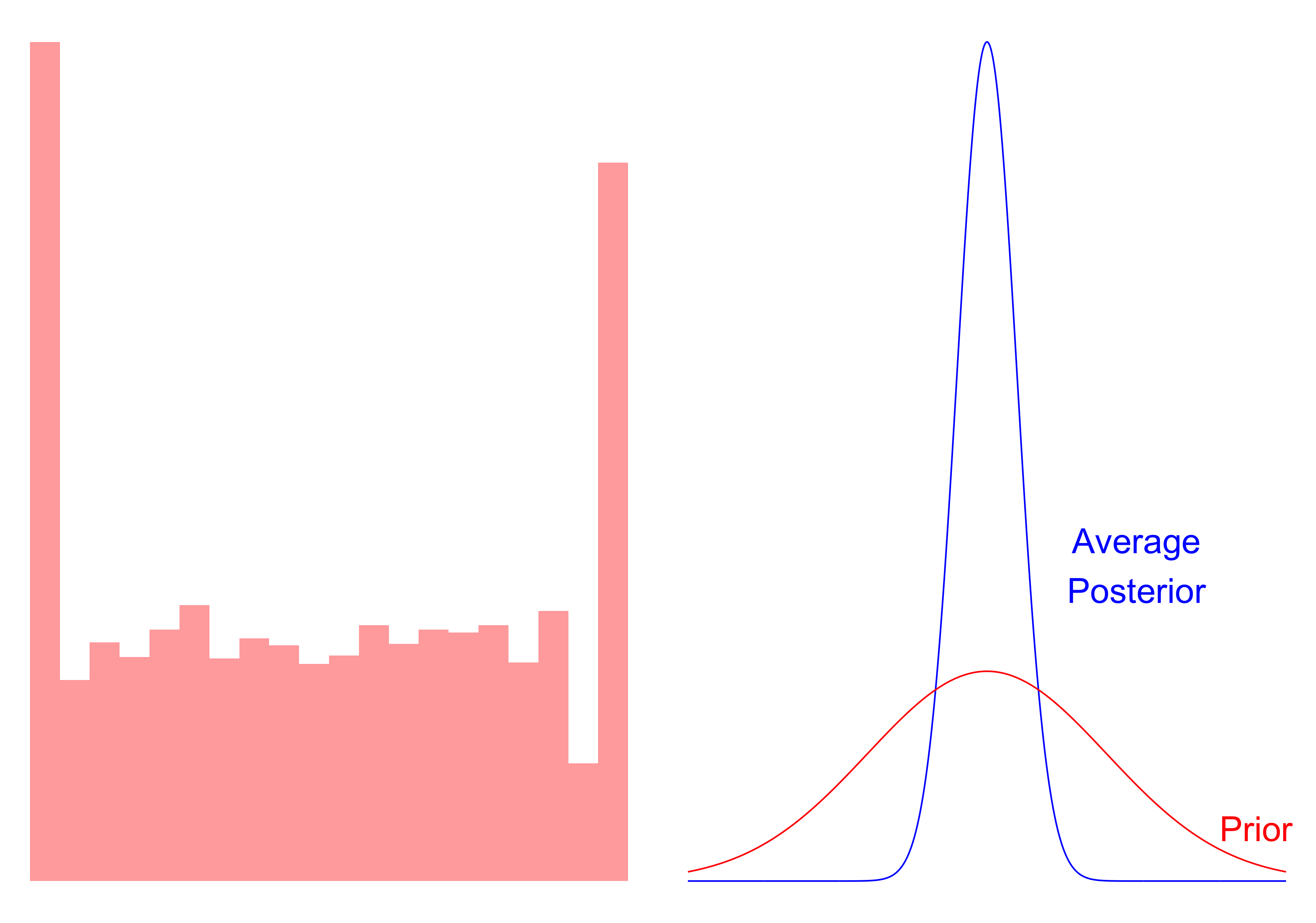}
                \caption{\textbf{Left panel}: Non uniform rank statistics with inflated frequencies on both ends (`U-shape' distribution). \textbf{Right panel}: Average posterior is underdispersed relative to the prior distribution. Non-uniformity in rank statistics comes from posteriors which are over-confident relative to the prior.}
                \label{fig:underdispersed}
            \end{figure}

            The shape of the distribution of rank statistics gives insight into how the MCMC may be miscalibrated. Specifically, it gives information about how the average posterior deviates from the prior distribution of a given parameter. Figures \ref{fig:underestimation} and \ref{fig:underdispersed} are recreated from \citet{talts2020validating} for convenience. These display examples of posterior miscalibration and the corresponding non-uniform rank statistics.

            The left panel of Figure \ref{fig:underestimation} shows a distribution of rank statistics with inflated frequency on the right side of the histogram. This distribution is consistent with posterior samples that tend to underestimate the prior distribution on average, as shown on the right panel. The sum of indicator random variables (see equation (13)) is large since a large proportion of posterior draws are smaller than the true value, resulting in a large rank statistic and hence a larger relative frequency on the right side of the histogram. The converse is true if the inflated frequency of the histogram is on the left hand side (i.e the average posterior over the SBC iterations overestimates prior). 

            The left panel of Figure \ref{fig:underdispersed} has large inflated frequencies on both ends of the histogram resulting in a `U-shape'. \citet{talts2020validating} describe this as under-dispersion of the posterior relative to the prior distribution on average. The average posterior is too narrow (or overconfident) relative to the spread of the prior, as seen on the right panel, resulting in inflated frequency on both ends of the rank statistic distribution. 
        
            \subsubsection{Chi-squared statistics}
            One difficulty of relying on the visualisation of rank statistic distributions for SV models is there are a very large number of unknowns in the model to consider. An alternative to visually checking for uniformity of a given parameter is to calculate a chi-squared statistic based on the counts in each histogram bin. Suppose the histogram has $J$ bins, where for each $j^{th}$ bin, $b_j$ is the number of counts and $e_j$ the expected count in bin $j$. The expected count is the ratio of $K$ rank statistics and $J$ bins, $e_j=\frac{K}{J}$. The corresponding chi-squared statistic is denoted by
            \begin{align}
            \chi^2 = \sum_{j=1}^J \frac{(b_{j} - e_{j})^2}{e_j}.
            \end{align}
            
            A discrete empirical distribution with the same number of realisations in each of $J$ bins will return a chi-squared statistic of zero, since the number of realisations in each bin is equal to the expected count under the relevant  uniform distribution. The distribution of chi-squared statistics is visualised to compare results across multiple simulations. Thus, the distribution of chi-squared statistics produced from the histograms of SBC rank statistics for a large number of univariate model parameters will give a high level summary of the overall model calibration, and hence for the overall performance of the algorithm. By comparing the distributions of chi-squared statistics produced by competing algorithms, one algorithm's degree of calibration, relative to that of the other algorithm, may be assessed. 

            \subsubsection{Effective sample size}
            Another univariate summary of a marginal posterior distribution is its effective sample size (ESS), which provides a measure of efficiency of the MCMC sampler. ESS estimates the number of effectively independent draws generated by a MCMC to estimate a target parameter. A poor ESS generally arises from high autocorrelation in the Markov chain, and thus is an indication of a highly dependent posterior sample. A more efficient MCMC algorithm will take a relatively lower level of resources (for example, time and number of draws) to get a sample from the target distribution when compared to a less efficient algorithm. If a MCMC algorithm possesses higher ESS for the majority of its parameters (relative to another strategy), this is seen as evidence of greater efficiency.
            
            The ESS is denoted as $n_{\text{eff}}$ and is given by
            \begin{align}
                n_{\text{eff}} = \frac{NM}{1+2 \sum_{l=1}^L \hat{\rho}_l}.
            \end{align}
            When more than one chain is generated from the target posterior, then the ESS quantity is a function of the number of chains $M$, the number of samples per chain $N$ and the autocorrelation at lag $l$, $\rho_l$ \citep{vehtari2021rank}. $L$ is a truncation lag chosen such that the sum of any two consecutive autocorrelation lags is negative \citep{geyer1992practical}. This truncation (or some kind of down-weighting) is required because a large number of lags leads to noisier autocorrelation estimates. Note that other approaches to estimating the true autocorrelation of the Markov chain and the ESS can be found in the literature. Details about this ESS formula, autocorrelation and other ESS calculations can be found in \citet{vehtari2021rank} and \citet{geyer1992practical}.

\section{MCMC Strategies}
This section defines and compares the two MCMC algorithms used to estimate the SV model. Section 4.1 and Section 4.2 describes Stan's No-U-Turn sampler and KSC's off-set mixture MCMC respectively. Section 4.3 compares the similarities and differences between both methods. Section 4.4 describes model reparameterisation in the context of each MCMC algorithm. 
    \subsection{Sampling method 1: Stan's Hamiltonian Monte Carlo with No-U-Turn Sampler}
        Hamiltonian Monte Carlo (HMC) is a general MCMC algorithm used to efficiently sample posterior draws associated with high-dimensional and complex stochastic models. Hamiltonian Monte Carlo, originally called Hybrid Monte Carlo, was developed in the physics literature \citep{duane1987hybrid} before being applied in the statistics literature by Radford Neal through his works in Bayesian Neural Networks \citep{neal1995bayesian} and statistical computing \citep{neal2011mcmc}. Versions of this algorithm have since become widely available through open source development projects, including Stan \citep{stan} and PyMC \citep{pymc2023}.

        The key innovation of HMC is that it uses the gradients of the target posterior distribution to generate an efficient path for the sampler to explore. HMC is able to reach more distant points with higher acceptance probabilities than many bespoke MCMC algorithms since its  proposal draws are made using information about the target distribution. Random walk samplers (such as Random Walk Metropolis Hastings) tend to be inefficient in higher dimensions since it becomes more difficult to randomly select a point with a high acceptance probability. The increasing complexity of high dimensions make it harder for a random walk sampler to explore regions over long distances, and thus the sampler tends to become stuck and will require many iterations to converge to the target distribution. 

        A thorough conceptual and theoretical explanation of HMC can be found in \citet{gelman2013bayesian} and \citet{betancourt2017conceptual}. HMC builds upon the Metropolis Hastings algorithm through the introduction of Hamiltonian equations, auxiliary momentum variables, and the gradients of the target log posterior distribution. The vector of model parameters $\theta$ is jointly sampled with a vector of momentum variables $\alpha$. The momentum variables are also called auxiliary variables since they are required as part of the Hamiltonian equations to generate an efficient proposal but are not a quantity of interest for inference.

        The sampled momentum variable sets out the proposal path for $\theta$ governed by the Hamiltonian dynamics (represented by a set of differential equations). These differential equations are solved using a discrete approximation algorithm, such as a leapfrog integrator, and is a function of the log posterior gradients for a given point $\theta$. The leapfrog integrator has a set of tuning parameters which determines the size and number of steps to be taken along the Hamiltonian path with the final step being the proposal value $(\theta^{\ast}, \alpha^{\ast})$. Finally, the Metropolis Hastings accept/reject step is applied using the parameters at the beginning of the leapfrog process $(\theta^{b-1}, \alpha^{b-1})$ and the proposal values $(\theta^{\ast}, \alpha^{\ast})$. A formal description of the HMC algorithm is provided in Appendix C.

        \subsubsection{Implementation}
        The Stan programming language's implementation of Hamiltonian Monte Carlo will be used for this study. Stan's default algorithm, the No-U-Turn Sampler \citep{hoffman2014no}, allows for direct sampling of the specified stochastic volatility model. The No-U-Turn Sampler (NUTS) proposes the final parameter using multinomial sampling biased towards the second half of the trajectory steps instead of applying the Metropolis-Hastings step \citep{betancourt2016identifying}. Stan samples the generative model directly and can flexibly handle complicated likelihood functions. 
        
        This approach will also use the same priors as specified in the KSC's off-set mixture approximation so that HMC is sampling from the same model (although other priors could be chosen).

    \subsection{Sampling method 2: Off-set Mixture MCMC}
        \citet{kim1998stochastic} sample the conditional posteriors of the stochastic volatility model using a mix of conjugate posterior distributions, Metropolis Hastings \citep{metropolis1953equation, hastings1970monte} within Gibbs \citep{geman1984stochastic} algorithms and an implementation of the Kalman Filter \citep{kalman1960new} and simulation smoother used to sample from the conditional posterior distribution of the latent states \citep{dejong1995}.\footnote{In this research the exact software to apply the simulation smoother is unavailable. So a more recent simulation smoother is used from the statsmodels Python package by \citet{seabold2010statsmodels}. The simulation smoother software within the statsmodels package is based on \citet{durbin2012time}. Koopman who co-authored \citet{durbin2012time} also co-authored the software used by \citet{kim1998stochastic} which can be found at \citet{koopman1996ssfpack}.} Note that for the simulation smoother to produce an appropriate posterior sample draw from the latent state vector, the state and measurement equations are required to be linear and conditionally Gaussian. Since the relationship between $y_t$ and $h_t$ in the measurement equation is not linear, a transformation is applied by squaring, adding an off-set, and taking the log of $y_t$, resulting in
        \begin{align}
        y_t^{*} &= \log(y_t^2 + c) = h_t + z_t, \label{eqn:transform}
        \end{align}
        where $z_t = \log(\epsilon_t^2)$ has a distribution that is like that of the logarithm of a chi-squared random variable. It is known that the error $z_t$ has mean -1.2704 and variance 4.93. The off-set $c$ is set to $c=0.001$ to ensure the model is robust to small $y_t^2$ values \citep{fuller1996introduction}. By transforming the entire measurement equation, as in (\ref{eqn:transform}), the relationship between $y_t^{*}$ and $h_t$ is now linear; however, the error is not Gaussian. Since it is not simple to sample from this parameterisation of the model, KSC use a mixture of seven Gaussian distributions to approximate the distribution of $z_t$, with the mixture selected to match the first 4 moments of the log chi-squared distribution. The mixture distribution is defined by
        \begin{align}
        f(z_t) = \sum_{i=1}^{I} q_if_N(z_t|m_i-1.2704, \nu_i^2),
        \end{align}
        where $K=7$, $f_N(z_t|m_i-1.2704, \nu_i^2)$ denotes the density of the normal distribution with mean $m_i-1.2704$ and variance $v_i^2$, and  component probabilities $q_i$, for $i=1,2,\ldots, I=7$.  The values of these hyperparameters and component weights are applied in this research and reported in Appendix A.

        Given a set of $T$ mixture component indicator variables, which are added to the MCMC scheme in order to use the approximating Gaussian mixture error distribution, the latent state vector may be sampled jointly via the Kalman Filter and simulation smoother, since the model is now conditionally linear and Gaussian. The static parameters $\mu$ and $\sigma^2$ are sampled directly from their conjugate posterior distributions whereas $\phi$ is sampled via a Metropolis Hastings accept/reject procedure.\footnote{The KSC method was a big advance over the one-state-at-a-time sampling method that was used prior to that time. There have also been other advances with particle MCMC since then that still exploit the insights gained from this method.}

        Since the mixture of Gaussians only approximates the distribution of $\log \epsilon_t$ up to its first four moments, \citet{kim1998stochastic} also recommend using a post-sampling importance sampling step to correct for approximation error. The importance weights used in this step are also applied in the context of SBC and can be found in Section 5.3. 
        
        \subsubsection*{Implementation}
        The code used to replicate the KSC algorithm\footnote{The KSC algorithm, KSC MCMC and `off-set mixture MCMC' will be used interchangeably. The model itself will be referred to as `off-set mixture model'.} was written by \citet{chad2018} and amended for the purposes of this research. Additional code was developed to produce rank statistics, quantities of interest as well as to integrate the sampling method into the SBC framework. Details of the full sampling algorithm are found in Appendix B.

    \subsection{MCMC methods comparison}
        HMC and the KSC sampling method are both MCMC strategies to generate samples from the target joint posterior distribution of the stochastic volatility model. However, the two sampling approaches are conceptually different in their design and implementation, despite sharing the same objective.

        HMC is a general approach to applying MCMC. The implementation of HMC is through the Stan programming language which allows for the specification of other generative or Bayesian models, not just stochastic volatility. The implementation details of HMC are abstracted away from the user, who only needs to focus on specifying the model using the Stan syntax. Any hyper-parameter tuning and initialisation is automated. Furthermore, Stan and HMC allows for the specification of different priors, whereas KSC's bespoke algorithm is dependent on the use of conjugate priors. 
        
        The KSC sampling method on the other hand is a bespoke algorithm designed specifically for the sampling of the stochastic volatility model. It is a strategy built around how to sample the parameters from the state space model in this specific context. The implementation of this MCMC requires careful attention to the sampling of each parameter using different tools and requires these components to be programmed manually, such as the Metropolis Hastings step and conjugate posterior distributions. 

        The conceptual difference in both methods result in different trade-offs. The abstraction of the MCMC implementation using the Stan programming language means users can focus more on the modelling and the applied research question. The algorithm is maintained and thoroughly tested by developers and different models can be compared with each other without worrying about differences in MCMC implementations. However, the abstraction away from the details means there is less flexibility in MCMC design in the scenario where the current instance of Stan or HMC is insufficient for a specific model.
        
    \subsection{Model Parameterisation}

        The parameterisation of a model can affect the performance of a MCMC algorithm when sampling from models with complex posterior geometries. An example of this is Neal's funnel \citep{neal2003slice} where the Hamiltonian Monte Carlo sampler encounters performance issues in hierarchical models and produces biased samples. Efficiency improvements of MCMC algorithms for state space models under different parameterisations are also explored in \citet{strickland2008parameterisation}.

        The stochastic volatility model as described in section 2.1 follows a centered parameterisation. This describes the central location of the latent state distribution which is centered on the mean of the log variance $\mu$. A model can be reparameterised in a way that might improve MCMC performance. Reparameterisation applies a transformation to the model which may change the interpretation of the parameters or the way they are sampled but remains mathematically equivalent to the original model. The reparameterisation available for a model is dependent on the MCMC approach. Some reparameterisation are available only to specific sampling strategies and may improve or impair a MCMC performance. 

        This research compares two different parameterisations for the two different sampling methods in a 2x2 design using SBC. This is summarised in Table \ref{tab:params}.
        \begin{table}
            \centering
            \begin{tabular}{|c|c|c|} \hline 
                 &  \textbf{Centered}& \textbf{Reparameterised}\\ \hline 
                 \textbf{HMC}&  Centered HMC& Reparameterised HMC\\ \hline 
                 \textbf{Off-set mixture MCMC (KSC)}&  Centered KSC& Reparameterised (Non-Centered) KSC\\ \hline
            \end{tabular}
            \caption{MCMC and parameterisation matrix}
            \label{tab:params}
        \end{table}
        
        \subsubsection{Reparameterised stochastic volatility in Stan}
        The reparameterisation for Hamiltonian Monte Carlo changes how the latent states are sampled. Specifically, the states are first sampled from a standard normal distribution and are transformed to be samples from the state equation. There are two steps to this:
        \begin{enumerate}
            
        \item Sample from a standard normal distribution and multiply by the variance of the states. The state vector is sampled with mean centered on zero and variance equal to $\sigma_{\eta}^2$ as given by
        \begin{align}
        h_{std} \sim&\space \,\mathrm{N}(0,1) \\
        h =& \space \, h_{std} \times \sigma_{\eta}.
        \end{align}
        \item Apply the appropriate re-scaling to get samples from the log variance
        \begin{align}
        h_1 =& \space \,\frac{h_{std, 1}\times \sigma_{\eta}} {\sqrt{1 - \phi^2}} + \mu \\
        h_{t+1} =& \space \,h_{std, t+1}\times \sigma_{\eta} + \mu  + \phi(h_{t} - \mu),\space \, t\neq 1.
        \end{align}
        \end{enumerate}
        This returns the log variance as desired.
        
        \subsubsection{Non-centered Gaussian mixture off-set model}
        The reparameterised model for the Gaussian mixture off-set is expressed differently due to the Kalman Filter. Reparameterised HMC samples from the joint posterior directly with state vectors centered at zero and transformed to return the correct log volatility estimates. Applying the Kalman Filter requires the state equation to be a function of the state variable with the average log variance parameter $\mu$ entering the measurement equation.

        The following parameterisation is described as non centered in location. This follows the methodology outlined in \citet{strickland2008parameterisation}. Starting with the log chi-squared model
        \begin{align}
        y^{\ast}_t =& \space \, h_t + z_t \\
        h_{t+1} =& \space \, \mu +\phi(h_t - \mu) + \sigma_{\eta} \eta_t.
        \end{align}
        Reparameterise the state variable $h_t$ by subtracting the average log variance $\mu$. Let $g_t = \space \, h_t - \mu$. This leads to the state equation
        \begin{align}
        g_{t+1} =& \space \, \phi g_t + \sigma_{\eta}\eta_{t}.
        \end{align}
        Return average log variance into measurement equation and rewrite as a function of the non centered state equation
        \begin{align}
        y^{\ast}_t =& \space \, g_t + \mu + z_t \\
        g_{t+1} =& \space \, \phi g_t + \sigma_{\eta}\eta_{t},
        \end{align}
        where the initial state is drawn from 
        \begin{align}
            g_1 \sim& \space \,\mathrm{N}\left(0, \frac{\sigma_{\eta}^2}{1-\phi^2}\right).
        \end{align}
        The mean of the log variance is now inside the measurement equation and demeaned from the state equation.
    
    \subsubsection{Prior predictive check}
        A prior predictive simulation of the second state parameter is performed for both a centered and reparameterised model using the approach described in Section 4.4.1. This is to demonstrate that different parameterisations express the same underlying mathematical model. This is similar to the first two steps of SBC, except instead of generating a data set, it generates the state variable implied by the prior draw. Let $\boldsymbol{\theta}$ be a vector of static parameters. Simulate a draw from the joint prior
        \begin{align}
        \boldsymbol{\theta}^{sim} \sim \pi(\boldsymbol{\theta}).
        \end{align}

        Generate the second state parameter conditional on prior simulation and first state variable
        \begin{align}
        h_2^{sim} \sim \pi (h_2|h_1, \boldsymbol{\theta}^{sim}).
        \end{align}

\begin{figure}[h]
            \centering
            \includegraphics[scale=0.1]{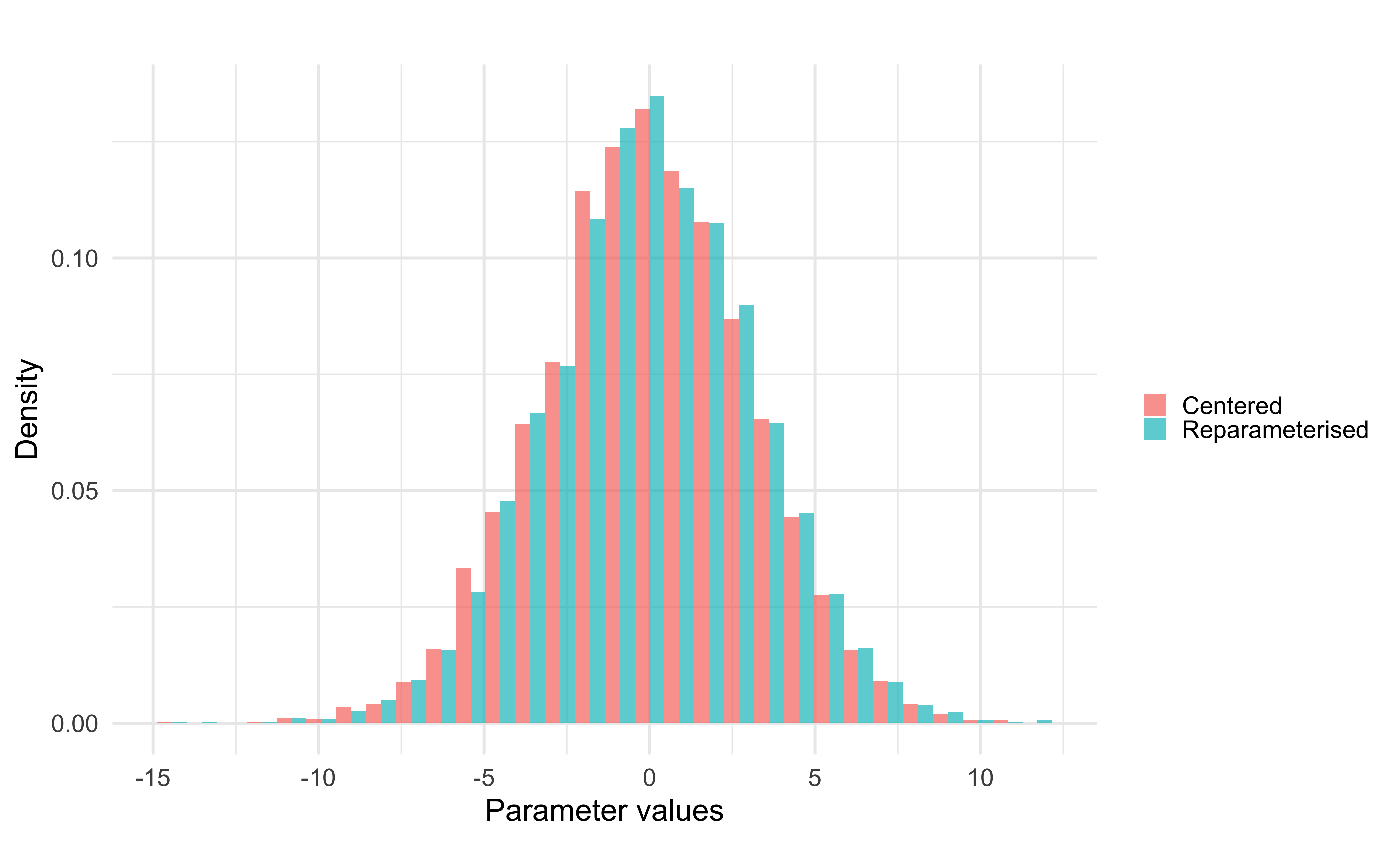}
            \caption{\textbf{Prior predictive simulation of second state variable from both centered and reparameterised stochastic volatility model}. Both models are mathematically equivalent and generate samples from the same data generating process.}
            \label{fig:priorpred}
        \end{figure}

        Simulating 5000 samples of $h_2^{sim}$ from both centered and non-centered models gives samples from the same data generating process as shown in Figure \ref{fig:priorpred}.

\section{Results}
    SBC is conducted for each MCMC sampling method and corresponding parameterisation as outlined in Table \ref{tab:params}. The priors in Section 2.1 are used for all simulations. For a given MCMC and parameterisation pair, 5000 SBC iterations\footnote{Results for 1000 SBC iterations returned noisy estimates and can be found in Appendix D.} are run using simulated data sets of size $T=1000$. 5000 rank statistics are obtained per univariate parameter from this process. 
    
    Rank statistic histograms are constructed with 20 bins for the static parameters and the 1st, 500th and 1000th latent state variables. The black horizontal line represents the expected counts in each bin if the algorithm is returning calibrated estimates and can be viewed as the uniform distribution. This is calculated to be 250 which is the ratio of  5000 SBC iterations and 20 bins. The chi-squared statistic is used to summarise the deviations from uniformity for all state variables since it is unrealistic to visually inspect all the histograms. The SBC algorithm is summarised in Algorithm \ref{alg:sbc}.

    Section 5.1 and Section 5.2 interpret the SBC results for HMC and KSC's MCMC method respectively. Section 5.3 describes the importance weights used to correct for approximation error when using the off-set mixture model and applying these weights to rank statistics. Section 5.4 presents a summary of rank statistic distributions for all latent state variables using chi-squared statistics. 
    
    \begin{algorithm}
        \caption{SBC}\label{alg:sbc}
        \begin{algorithmic}
        \For{\texttt{k in} $1:5000$}
                \State \text{Draw from joint prior: } $\boldsymbol{\theta}^{sim}_k \sim\pi (\boldsymbol{\theta})$
                \State \text{Simulate data set with 1000 observations: } $\boldsymbol{y}^{sim}_k \sim \pi(\boldsymbol{y}|\boldsymbol{\theta}^{sim}_k)$
                \State \text{Draw 999 posterior samples post burn in:} $\{\boldsymbol{\theta}_1,\dots , \boldsymbol{\theta}_{999}\}_k \sim \pi(\boldsymbol{\theta} | \boldsymbol{y}^{sim}_k)$
                \State \text{Compute rank statistics:} $\boldsymbol{r} = \mathrm{rank}(\{\boldsymbol{\theta}_1,\dots , \boldsymbol{\theta}_{999}\}_k, \boldsymbol{\theta}^{sim}_k)$
              \EndFor
        \end{algorithmic}
        \end{algorithm}
        
    \subsection{Hamiltonian Monte Carlo (No-U-Turn Sampler)}
\begin{figure}[H]
        \centering
        \includegraphics[scale=0.09]{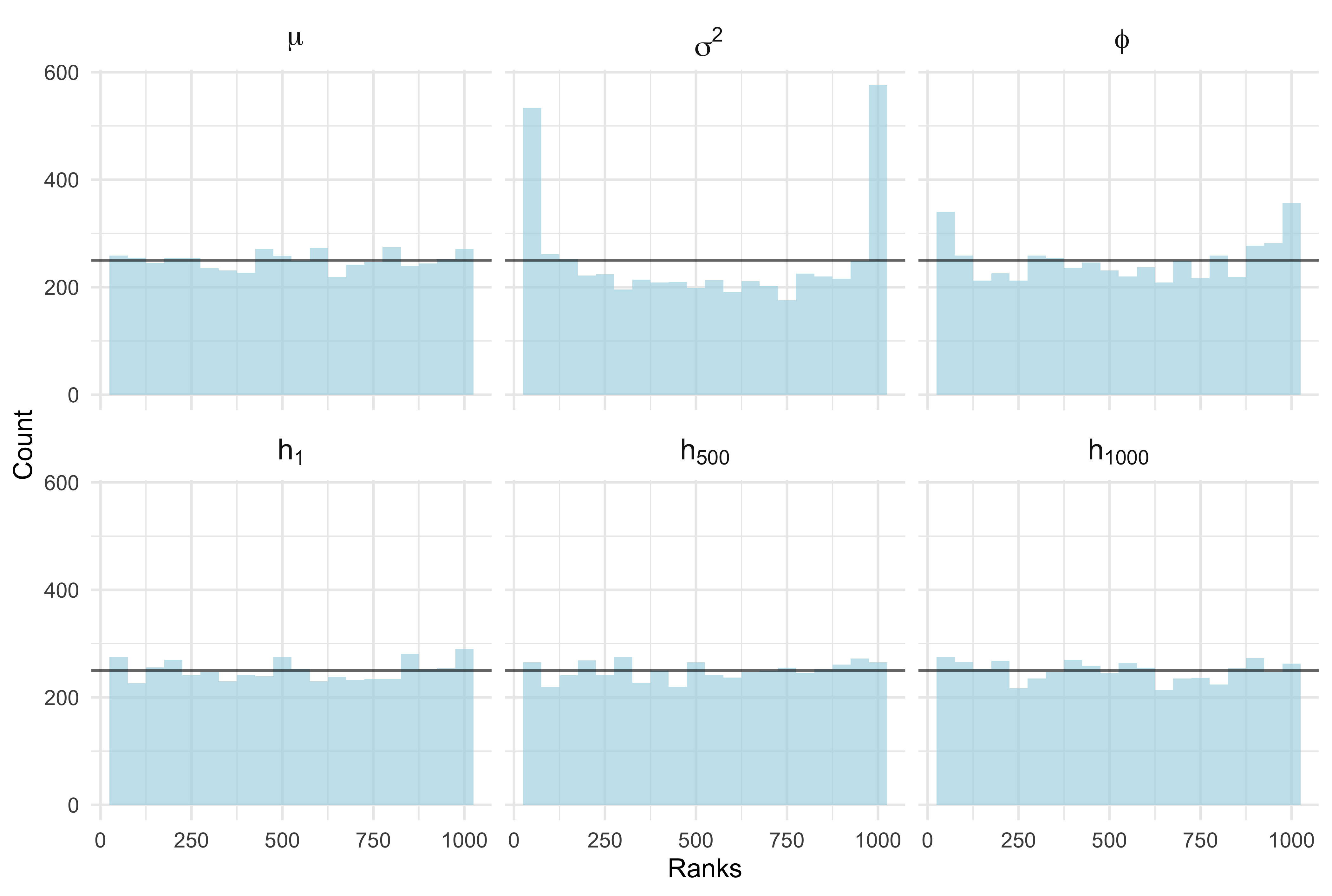}
        \caption{\textbf{Distribution of rank statistics with 5000 SBC iterations for centered model using Hamiltonian Monte Carlo}. $\mu$ and the state parameters look approximately uniform. $\sigma^2$ still has inflated frequencies on both the left and right hand side indicating under-dispersion relative to the prior distribution. $\phi$ exhibits a similar U-shape to $\sigma^2$ although the U-shape is much smaller.}
        \label{fig:cphmc5k}
    \end{figure}

    SBC results for the centered model using HMC are displayed in Figure \ref{fig:cphmc5k}. The centered parameterisation displays a lack of uniformity in rank statistics for $\sigma^2$ and $\phi$. The $\sigma^2$ parameter displays a U-shaped histogram with inflated frequencies on both ends while $\phi$ exhibits a U-shape with relatively smaller inflated frequencies. This indicates the MCMC algorithm is not returning calibrated estimates for these parameters. The average posterior samples of $\sigma^2$ and $\phi$ over the SBC iterations are \textit{under-dispersed}, or \textit{overconfident}, relative to the prior distribution. This implies the posterior geometry of these parameters may be difficult to sample for the HMC algorithm.

\begin{figure}[H]
        \centering
        \includegraphics[scale=0.09]{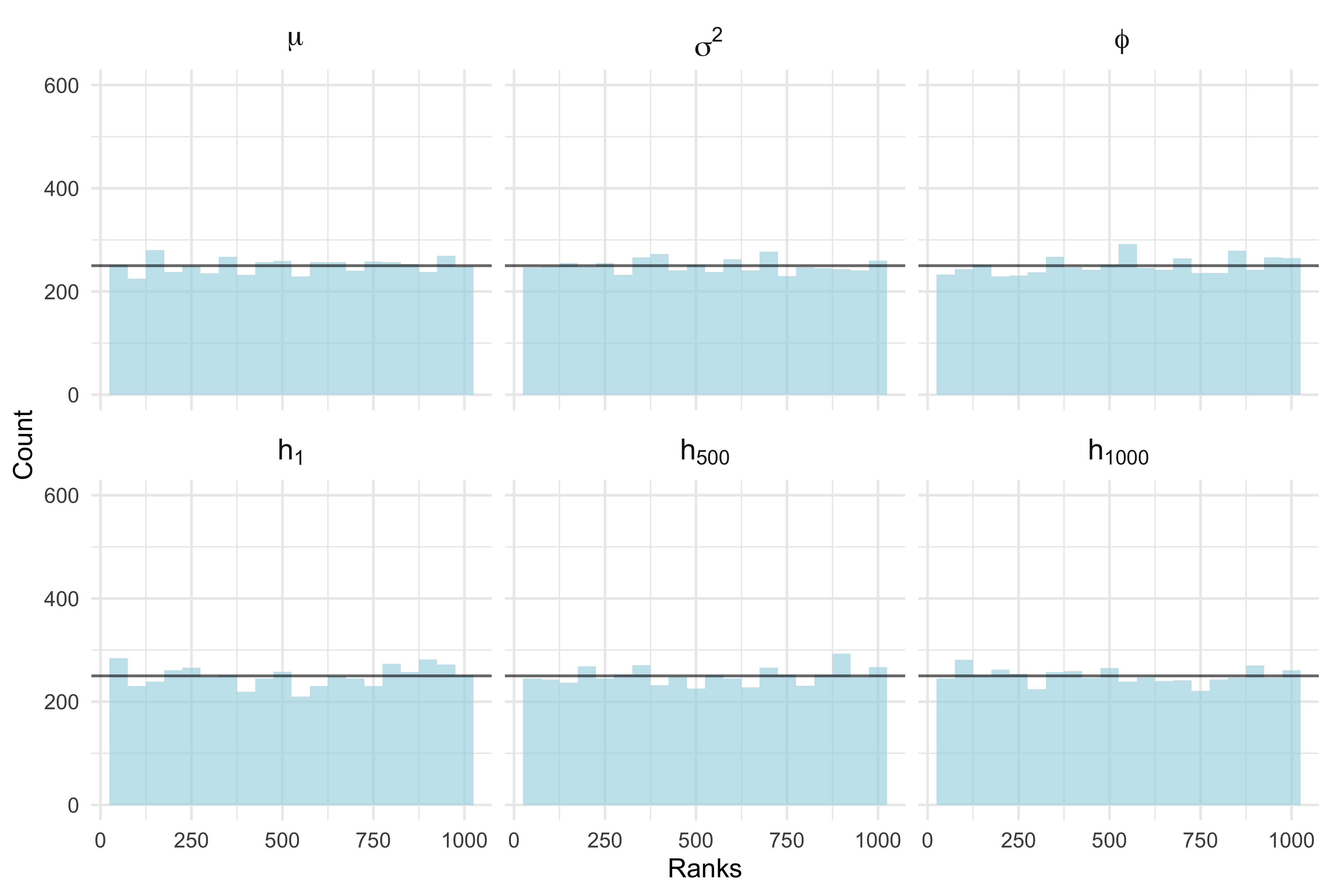}
        \caption{\textbf{Distribution of rank statistics with 5000 SBC iterations for reparameterised model using Hamiltonian Monte Carlo}. The rank statistic distributions for $\sigma^2$ and $\phi$ appear to follow a uniform distribution. This suggests that the HMC algorithm is returning the correct posterior estimates on average for the displayed parameters.}
        \label{fig:ncphmc5k}
    \end{figure}

    Figure \ref{fig:ncphmc5k} show the SBC results for the reparameterised model using HMC. The reparameterised model produces approximately uniform rank statistics across all reported parameters. The distributions of $\sigma^2$ and $\phi$ now exhibit a more uniform shape, implying greater calibration and correct posterior estimates returned on average. These results suggest that HMC struggles to sample some parameters from the centered parameterisation of the stochastic volatility model. The reparameterised model possess stronger evidence of calibration for the parameters examined, producing more reliable and consistent inference.

\begin{figure}[H]
        \centering
        \includegraphics[scale=0.09]{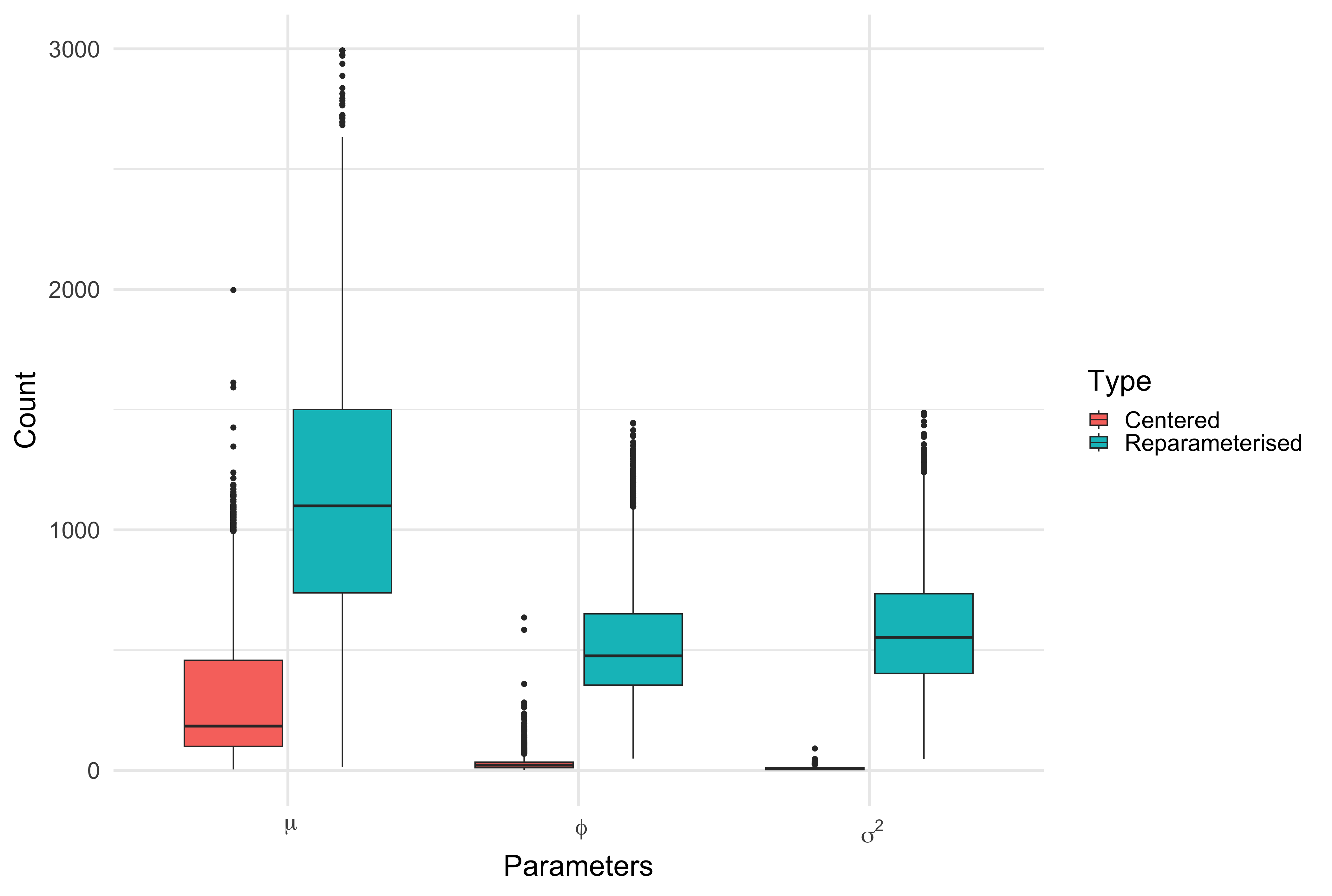}
        \caption{\textbf{Effective sample sizes for static parameters after 5000 SBC iterations}. The centered parameterisation struggles to generate independent samples for $\sigma^2$ and $\phi$. The reparameterised model demonstrates much more efficient sampling behaviour.}
        \label{fig:hmcess}
    \end{figure}

    Computational difficulty in the centered parameterisation is reflected in the effective sample sizes. Figure \ref{fig:hmcess} and Table \ref{tab:hmcess} summarise the ESS distributions for the static parameters in both parameterisations after 5000 SBC iterations. The reparameterised model generates independent samples much more efficiently than the centered model, reflected in the larger ESS values across every summary statistic. This is consistent with HMC struggling to sample from the posteriors of the static parameters overall in the centered model. 

    \begin{table}[H]
        \centering
        \begin{tabular}{|c|c|c|c|c|c|c|c|} \hline 
        Parameter &  Type&Min& q25&  Median& Mean & q75&Max\\ \hline 
        $\mu$&  Centered&3.79 & 99.7 & 184. & 305. & 457. & 1997.  \\
     $\mu$&  Reparam&14.5 & 738. & 1099. & 1143. & 1500. & 2993.  \\\hline 
     $\phi$&  Centered&1.44 & 11.0 & 21.7 & 26.3 & 34.1 & 635.  \\
     $\phi$&  Reparam&48.8 & 355. & 476. & 521. & 651. & 1445.   \\ \hline 
     $\sigma^2$&  Centered& 1.17 & 2.99 & 7.04 & 8.05 & 11.3 & 90.7 \\ 
     $\sigma^2$&  Reparam&46.1 & 403. & 553. & 585. & 734. & 1486. \\ \hline
        \end{tabular}
        \caption{\textbf{ESS for centered and reparameterised SV model across 5000 SBC iterations for HMC}. The reparameterised model has higher ESS counts for each summary statistic relative to the centered parameterisation.}
        \label{tab:hmcess}
    \end{table}

    \subsection{Off-set Mixture MCMC}
    The off-set mixture MCMC is run with the first 10,000 samples discarded as burn-in (following the method of \citet{kim1998stochastic}). However, unlike KSC who take 750,000 posterior draws from the model, the number of post burn-in posterior samples taken as part of the SBC procedure is 9,999. Taking 750,000 draws from the off-set mixture model would lead to issues due to disk space limitations and computational resources required to run SBC. Furthermore, the number of posterior samples is larger than the 999 taken for the SBC experiment run for Hamiltonian Monte Carlo. Earlier experiments attempted 999 post burn-in draws for the off-set mixture model, however there were concerns that such a short chain would not have converged onto the target posterior distribution. Therefore 9,999 draws was chosen as a compromise under both constraints. 

\begin{figure}[H]
        \centering
        \includegraphics[scale=0.09]{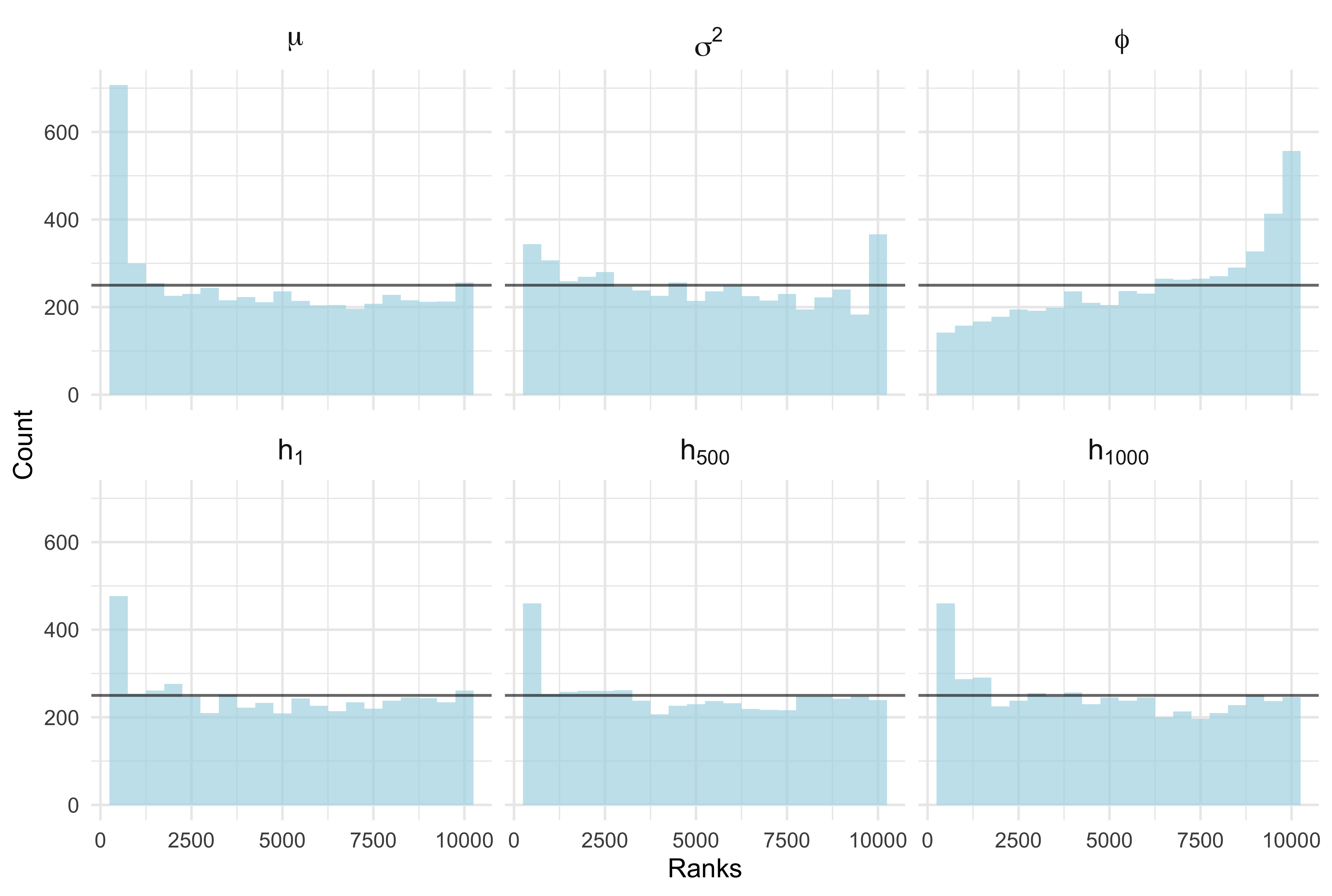}
        \caption{\textbf{Distribution of rank statistics with 5000 SBC iterations for centered Gaussian mixture approximation model}. The rank statistics for all the selected parameters display deviations from uniformity. Large inflated frequencies exist on the left-hand side for $\mu$ and the latent state variables and there is an inflated frequency on the right-hand side for $\phi$.}
        \label{fig:cpksc5k}
    \end{figure}

    Both the centered and the non-centered Gaussian off-set mixture model contain issues with calibration. Figure \ref{fig:cpksc5k} presents the results from the centered model. There is a lack of uniformity across all parameters. In particular, there is an inflated frequency on the left hand side for $\mu$ and the latent state variables, indicating that the posterior samples on average are overestimating the prior distribution. There is also an inflated frequency on the right hand hand side of $\phi$, implying underestimation of the prior by the average posterior over the SBC iterations. 

\begin{figure}[H]
        \centering
        \includegraphics[scale=0.09]{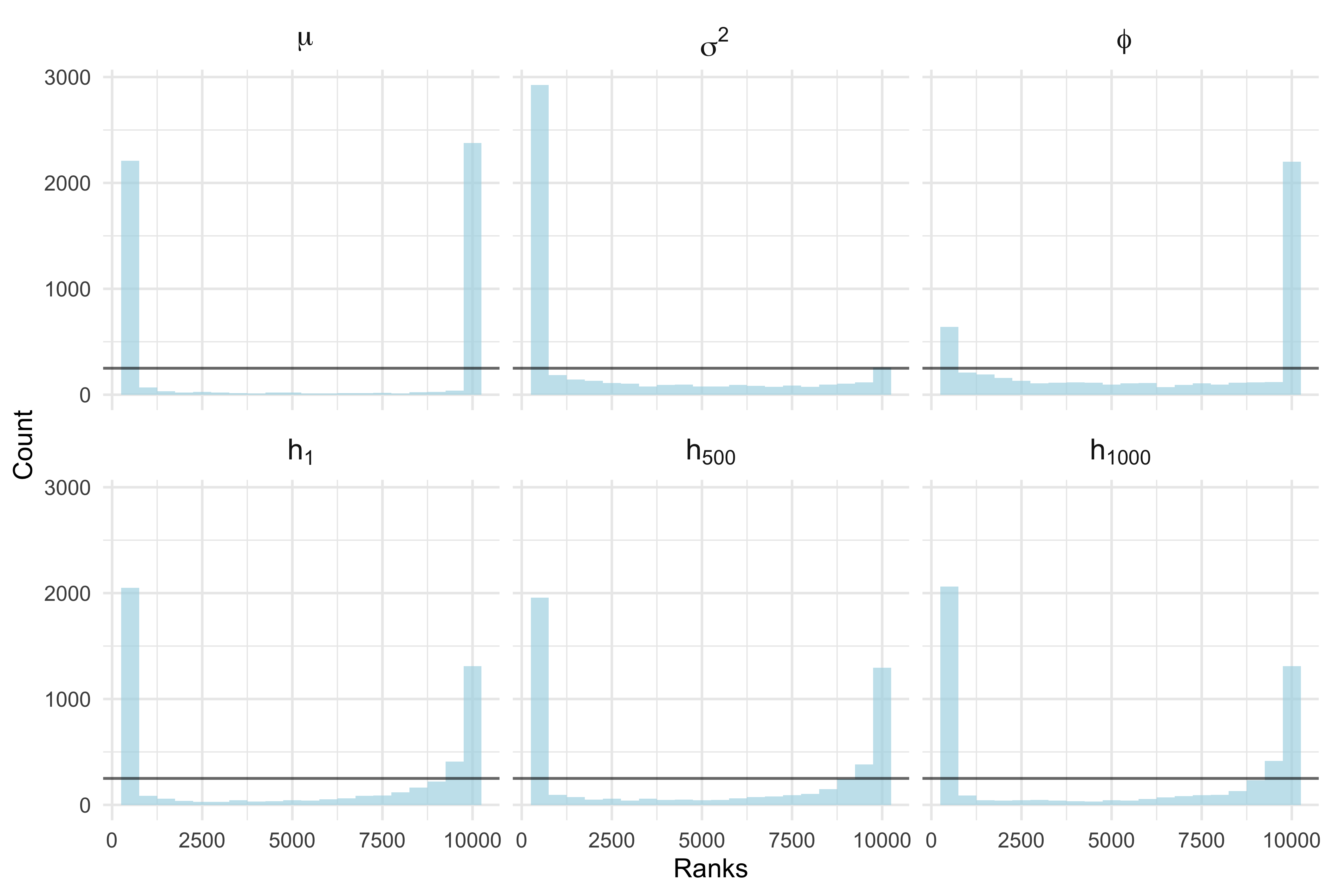}
        \caption{\textbf{Distribution of rank statistics with 5000 SBC iterations for non centered Gaussian mixture off-set model}. The rank statistics for all selected parameters are non uniform in shape. The KSC MCMC appears uncalibrated for both parameterisations of the model.}
        \label{fig:ncpksc5k}
    \end{figure}

    The non-centered model in location exhibits larger issues in calibration as shown in Figure \ref{fig:ncpksc5k}. Each histogram exhibits large inflated frequencies at one or both ends, suggesting major issues in the sampler returning correct posterior estimates. The KSC MCMC algorithm applied to this parameterisation of the model looks uncalibrated overall. The results of both parameterisations suggest applying this MCMC algorithm on real data will on balance produce incorrect and unreliable posterior samples on average (conditional on the model and the number of post burn-in draws).

\begin{figure}[H]
        \centering
        \includegraphics[scale=0.1]{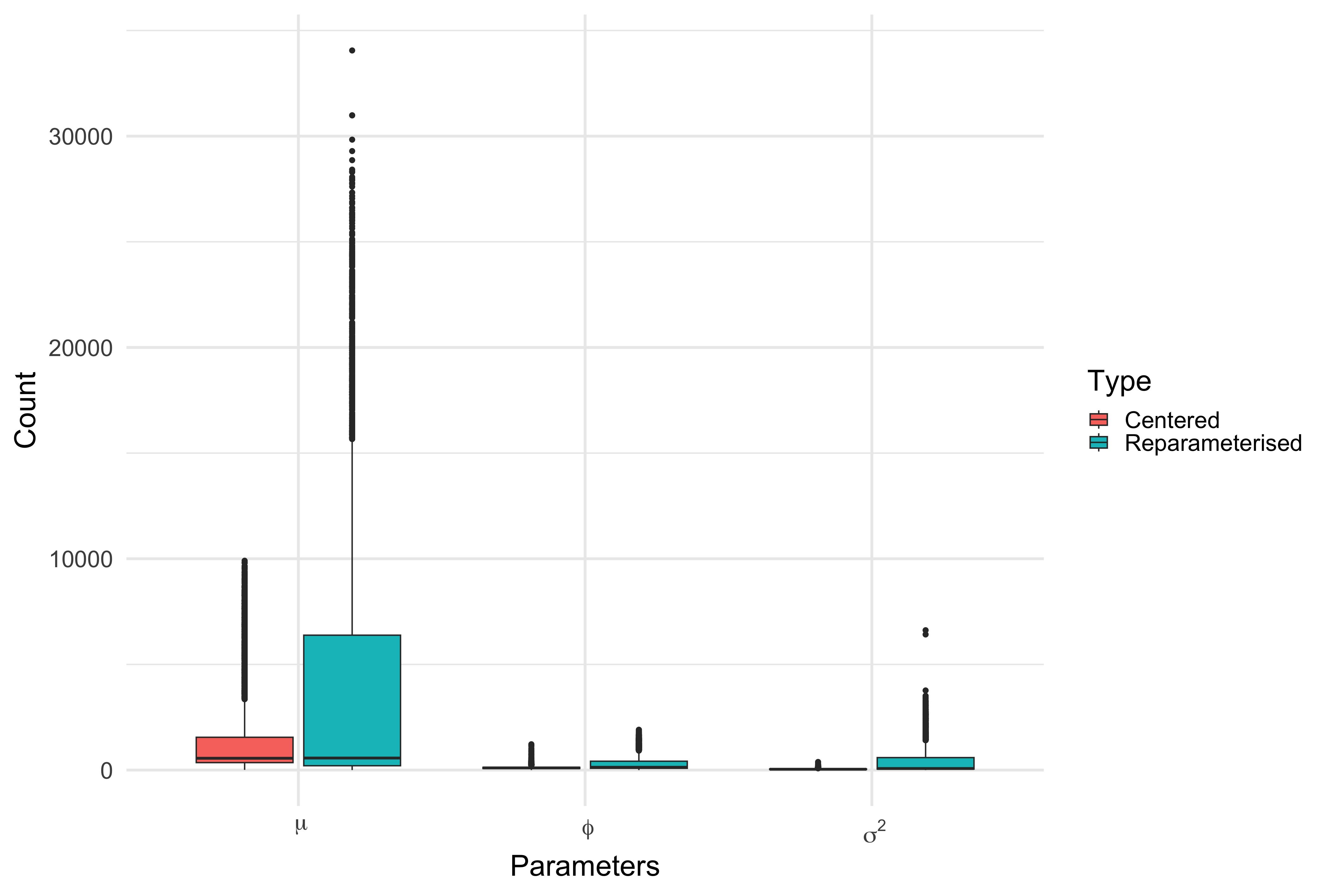}
        \caption{\textbf{Effective sample sizes for static parameters after 5000 iterations of SBC and 9999 post burn-in draws for the off-set mixture model}. The KSC sampling strategy struggles to generate independent samples for all static parameters and parameterisations, suggesting a high degree of autocorrelation in the Markov chain.}
        \label{fig:kscess}
    \end{figure}

    The ESS estimates for this model indicate a high degree of autocorrelation and difficulty in generating independent samples from both parameterisations. The numerical results are summarised in Figure \ref{fig:kscess} and Table \ref{tab:kscess}. The estimate for $\mu$ in the reparameterised model exhibits a long right tail, with values larger than the 9,999 post burn-in posterior draws suggesting the presence of negatively autocorrelated samples. The median ESS for each static parameter relative to the number of posterior draws is very small. There is some improvement for the reparameterised model with higher ESS values across some of the static parameter quantiles, but most of these improvements are negligible. Overall, the off-set mixture MCMC is inefficient at sampling the static parameters. Note, Figure \ref{fig:hmcess} and Figure \ref{fig:kscess} are not directly compared due to scale difference from varying posterior draws.

    \begin{table}[H]
        \centering
        \begin{tabular}{|c|c|c|c|c|c|c|c|} \hline 
        Parameter &  Type&Min& q25&  Median& Mean & q75&Max\\ \hline 
        $\mu$&  Centered&9.57 & 352. & 559. & 1482. & 1552. & 9907.\\
     $\mu$&  Reparam&1.99 & 205. & 569. & 4174. & 6384. & 34055.\\\hline 
     $\phi$&  Centered&2.47 & 74.0 & 103. & 112. & 130. & 1222.\\
     $\phi$&  Reparam&1.03 & 83.4 & 132. & 325. & 419. & 1907. \\ \hline 
     $\sigma^2$&  Centered&2.33 & 29.0 & 39.5 & 44.2 & 52.8 & 384. \\ 
     $\sigma^2$&  Reparam&1.25 & 47.4 & 73.4 & 399. & 591. & 6618. \\ \hline
        \end{tabular}
        \caption{\textbf{ESS for centered and reparameterised SV model across 5000 SBC iterations and 9999 post burn-in draws for the off-set mixture model}.}
        \label{tab:kscess}
    \end{table}

    The results from the off-set mixture algorithm indicates issues with calibration. The centered parameterisation appears to be more favourable for this sampling approach. Other ways of improving the sampling of this model using this MCMC strategy could be to use another parameterisation such as non centered in scale.

    There are a few potential reasons for the poor calibration results. It may be due to the ineffectiveness of the MCMC strategy to generate the correct posteriors. Additionally, it could be due to the approximation of the actual stochastic volatility model not producing accurate posterior estimates. Correcting approximation error of the model is explored in the next section.

    \subsection{Correcting approximation error using importance weights}
    \citet{kim1998stochastic} correct for approximation error in their method by using importance weights. The re-weighting procedure ensures that samples are drawn from the correct posterior density. This is applied to the calculation of the expected value of the stochastic volatility posterior density using samples drawn from the off-set mixture model. 

    Importance weights are defined as the ratio of the joint posterior from the SV model and the posterior distribution of the approximate model. A weight is produced for each posterior sample generated by MCMC. These weights correct the samples from the approximate distribution by increasing or decreasing the contribution of that sample in the calculation of the expectation. 

    These weights can be applied in an additional sampling step using Metropolis Hastings or Importance Re-sampling (also known as sampling-importance re-sampling or SIR \citep{gelman2013bayesian}) to produce samples from the target posterior. However, the rank statistic can also be written as a function of the weighted expectation of the indicator random variable. This re-weighting step is applied to the calculation of the rank statistics to see if the correction improves the calibration of the MCMC sampler.\footnote{Efficiency estimates of the re-weighted posterior samples are omitted. It was unclear at the time of writing whether the ESS could be written as a function of the expectation or whether the ESS from re-sampled posterior samples using the importance weights are valid.}

    \subsubsection{Re-weighting rank statistics of the Gaussian off-set mixture model}
    Let $v(\theta, h)$ be the log weights defined as the log difference between the posterior densities of the true model with log chi-squared errors $\log c(\theta, h | y^{\ast})$ and the off-set mixture model $\log d(\theta, h | y^{\ast})$. The is given in the following expression
    \begin{align}
        v(\theta, h) = \log c(\theta, h | y^{\ast}) - \log  d(\theta, h | y^{\ast}) = \text{const} + \log c(y^{\ast}|h) - \log d(y^{\ast} | h).
    \end{align}

    Taking the exponential and normalising the weights for the $b^{th}$ posterior draw (note the constants cancel out) returns
    \begin{align}
    w^b = \frac{exp(v(\theta, h)_b)}{\sum_z exp(v(\theta, h)_z)}
    \end{align}
    which gives the normalised importance weight. The expectation for any function of the posterior samples can be written as a function of these weights. Let $S(\theta)$ be an indicator random variable that is a function of the posterior samples. The expectation is given as
    \begin{align}
    S(\theta) &= 1[f(\theta_b) < f(\theta^{sim})] \\
    E[S(\theta) | y] &= \int S(\theta) c(\theta | y) d\theta\\ 
    &= \frac{\int S(\theta)\times exp(v(\theta, h)) * d(\theta, h | y^{\ast})d\theta d h}{\int exp(v(\theta, h)) * d(\theta, h | y^{\ast})d\theta d h}.
    \end{align}
    
    Therefore, the expectation of the re-weighted posterior samples can be expressed by
    \begin{align}
    E[S(\hat{\theta}) | y^{\ast}] = \sum_b^B S(\hat{\theta}_b)w_b.
    \end{align}

    The rank statistic can be rewritten as a function of the expectation and weights. This gives us the re-weighted rank statistics
    \begin{align}
    r = \sum_{b=1}^{B}1[f(\theta_{b}) < f(\theta^{sim})] \approx  B\times E[S(\hat{\theta})] = B\times \sum_b^B S(\hat{\theta}_b)w_b.
    \end{align}

\begin{figure}[H]
        \centering
        \includegraphics[scale=0.1]{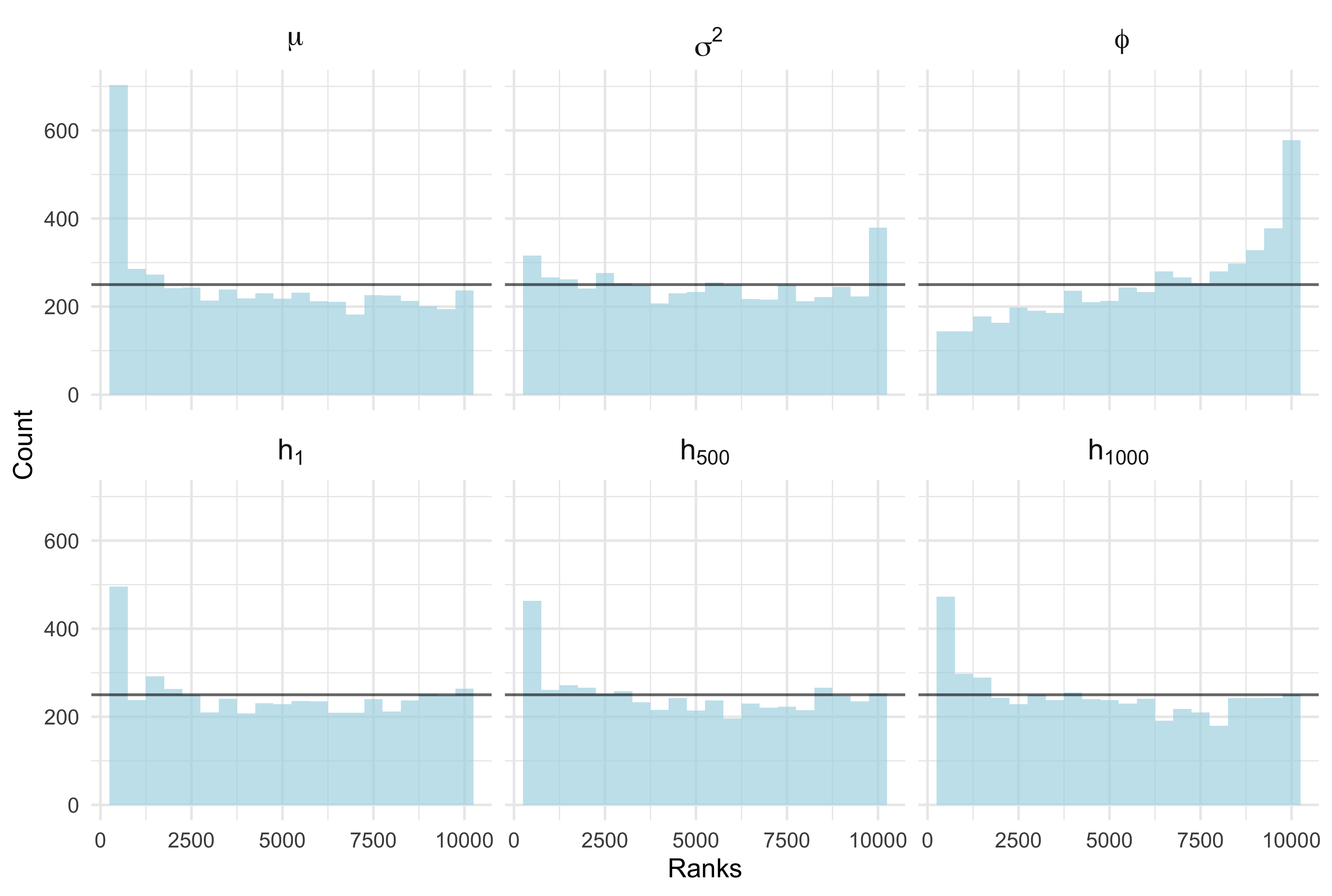}
        \caption{\textbf{Distribution of rank statistics for re-weighted rank statistics from the off-set mixture model}. The shape of the histogram is consistent with the unweighted rank statistics.}
        \label{fig:reweight5k}
    \end{figure}

    The results from applying this re-weighting step to the rank statistics of the centered model are given on Figure \ref{fig:reweight5k}. There are no improvements to the uniformity of the rank statistic distributions. The shape of the histograms is consistent with the shape of the unweighted rank statistics. Overall re-weighting the posterior samples from the approximate model does not improve the calibration results. The re-weighted rank statistics for the reparameterised model also did not improve and can be found in Appendix E.

    \subsection{Evaluating SBC for state variables}
    It is not practical to inspect the histograms of all parameters and states in the stochastic volatility model. Instead, the chi-squared statistic is used to summarise the shape of the rank statistic distribution. This will be used to summarise the state variables only since the static parameters can be compared across simulations individually. 

\begin{figure}[H]
        \centering
        \includegraphics[scale=0.1]{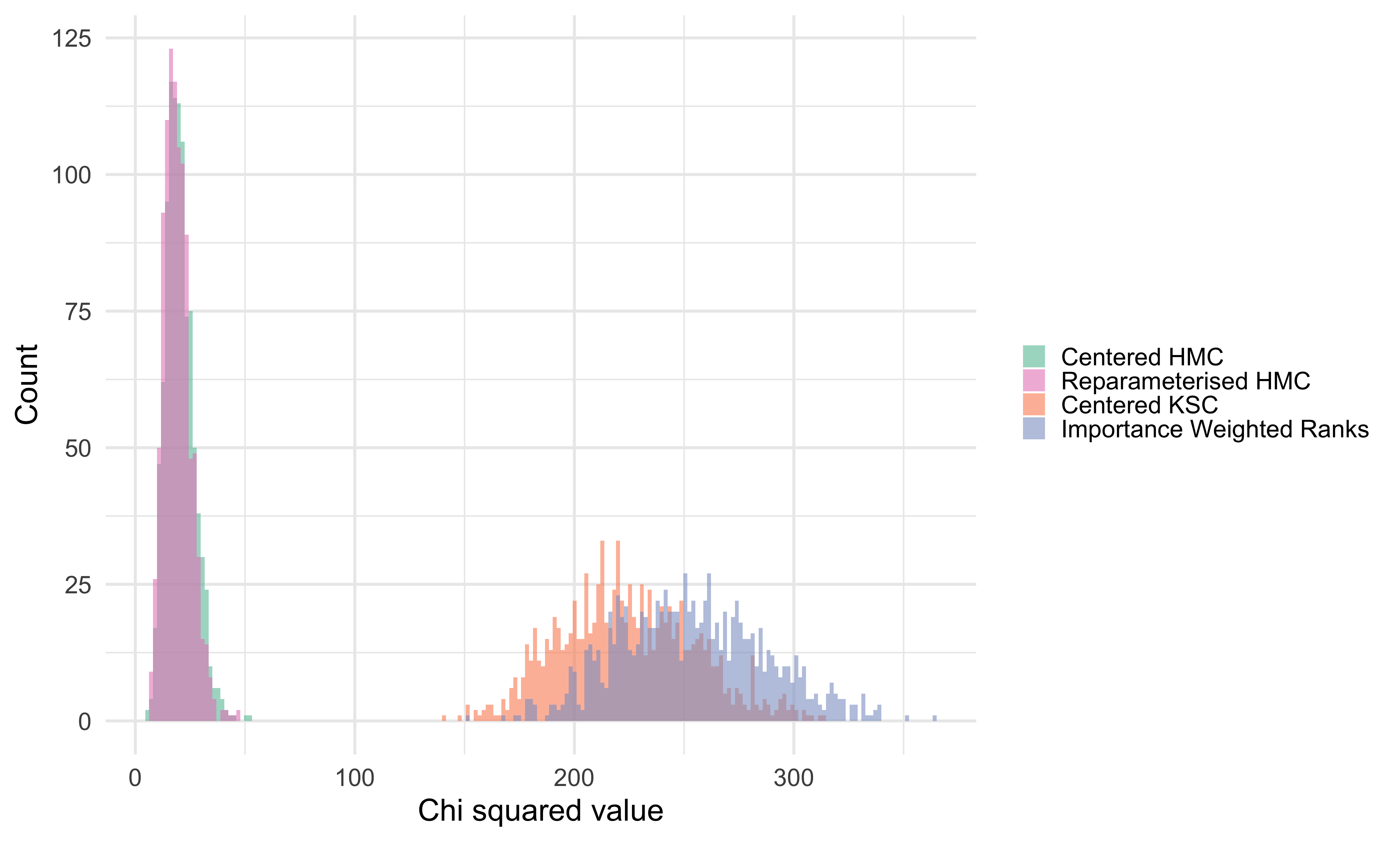}
        \caption{\textbf{Distribution of chi-squared statistics for latent state variables from 5000 SBC iterations}. Values closer to 0 are more uniformly distributed. Both HMC simulations are much closer to zero than KSC. This suggests that the HMC algorithm overall produces more calibrated posterior estimates for the latent state variables.}
        \label{fig:allchisq}
    \end{figure}

    The chi-squared statistics for centered and reparameterised HMC, centered KSC and centered importance weighted KSC are reported as histograms in Figure \ref{fig:allchisq} using results from 5000 SBC iterations. Reparameterised KSC is omitted as its poor SBC performance skews the scale of the plot which makes it difficult to analyse results. A chi-squared statistic of 0 means the rank statistics for a given parameter is exactly uniformly distributed and any value away from 0 captures deviations from uniformity. The chi-squared statistics for the latent state variable ranks are expected to be some non-zero value since they are calculated from finite samples and subject to sampling variation even if the draws come from a uniform distribution. The purpose of this plot is to provide a high level summary of the latent state variable calibration for each simulation - how close their rank statistics are to being uniformly distributed and their distance relative to the other algorithms and parameterisations. 
    
    The HMC results are relatively close to zero. Both KSC and the importance weighted ranks on the other hand are far from zero suggesting larger deviations from uniformity. There is a wide gap between the HMC and KSC chi-squared statistics implying the distribution of latent state rank statistics are much more  uniform for HMC relative to KSC. Combined with the results in the previous section, there is a strong evidence that the HMC algorithm outperforms the KSC algorithm where it comes to calibration across all parameters and latent state variables.

\section{Discussion}
This research applied a simulation study to evaluate the calibration and efficiency of different MCMC algorithms in the context of the stochastic volatility model. Simulation Based Calibration was used to determine whether these algorithms were returning the correct posterior estimates on average under different SV model parameterisations. 

Hamiltonian Monte Carlo applied on the reparameterised model performs best based on calibration and efficiency results. Centered parameterisation using HMC performed moderately well but struggled to produce calibrated posterior estimates for $\sigma^2$ and $\phi$ as well as generate independent samples for these parameters. The off-set mixture MCMC struggled to return calibrated posterior estimates and efficient samples for both parameterisations. However, the centered parameterisation results for the Gaussian mixture model may improve if we increased the length of the Markov chain. This is discussed later in the limitations section. 

These results demonstrate that the performance of a MCMC algorithm is sensitive to the parameterisation of the model in the context of calibration and efficiency. Furthermore, the parameterisation of a model is conditional on the choice of MCMC algorithm. Evaluating the effectiveness of a modelling strategy (i.e choice of model, parameterisation and sampler) on simulated data gives a lot of a priori information about how the model will perform before it is fit on real data. Performing SBC may help isolate confounding factors in computation if any problems occur. 

The priors chosen for this study came directly from \citet{kim1998stochastic}. The SBC results for the Gaussian off-set mixture model may improve if the priors were more informative - although this is only conjecture. Whether or not to use tighter priors depends on the modelling context and the problem at hand \citep{gelman2017prior}. If a model is expected to handle data generated by the priors, then un-calibrated results from SBC raises questions about the suitability about the model choice, MCMC or parameterisation. 

Other parameterisations of the stochastic volatility model are available as outlined in \citet{strickland2008parameterisation}, namely non centered in scale or both non centered in scale and location. The failure of non centered parameterisation in location for the off-set mixture model could be due to a variety of reasons. This may be due to a particular software implementation of the Kalman Filter and Simulation Smoother (where a specific software package was applied in this research) that was not designed to handle the non-centered parameterisation. Or perhaps this particular parameterisation does not suit the sampling strategy applied by KSC and another approach (e.g. HMC) might be more successful. Lots of different design and implementation details could be explored further. Pinpointing the reason why this model parameterisation failed as well as exploring other configurations and MCMC algorithms in this context is left for future research.

It is worth noting that there are many ways of comparing MCMC approaches. Other features that may be of interest is the speed in which a MCMC can converge onto the target distribution and generate samples for inference. However, speed is difficult to define and is subject to a variety of different factors such as hardware, operating system, and software versions. The purpose of this research is to compare algorithms based on calibration and efficiency. Additionally, this simulation study does not provide any advice on the appropriateness of the stochastic volatility model on real data. SBC implicitly assumes the data generating process is coming from the `correct model' and does not say which MCMC algorithm is more calibrated when the model is misspecified. Whether the stochastic volatility model is appropriate for modelling a particular financial time series requires further study into the data generating process, domain expertise and a suite of other diagnostic checks. Examples of this are posterior predictive checks and out of sample predictive performance. Rather, the SBC approach gives insight into how well calibrated a sampling strategy is conditional on a known data generating process which is important in all applied contexts.  

\subsection{Limitations}
A limitation to applying SBC in any modelling context is that it is computationally intensive. SBC requires fitting multiple models to get reliable estimates. In some cases, fitting just one model can take a long time, let alone multiple. Producing results for this research within a reasonable time-frame required use of a high performance cluster. Computing infrastructure may limit the opportunity for SBC to be performed. The choice of SBC iterations for this research was arbitrary and a more principled approach to picking the number of SBC iterations could be explored in other research.

Diagnostic tools for evaluating SBC results and models with many parameters is an area for improvement. As discussed, inspecting many histograms for highly parameterised models is unrealistic. Other summary statistics and visualisations can be applied for a high level comparison, such as the chi-squared statistics applied in this research or formal hypothesis tests of uniformity such as the chi-squared test. Furthermore, it is not always straightforward to understand why any one parameter may produce poor calibration results, for example, an arbitrary latent state variable. SBC may tell us some information about the miscalibration based on the histogram shape (e.g. average posteriors underestimating or overestimating the prior distribution, underdispersed or overdispersed average posteriors relative to the prior distribution), but understanding why this occurs may not be immediately clear. Although, diagnosing sampling problems with complicated posterior geometries speaks more generally to the difficulty of understanding complex models in the first place as opposed to a limitation of SBC.

Lastly, it is unclear what constitutes a fair comparison between algorithms when comparing SBC results. In particular, the number of post burn-in or warm-up samples differed between algorithms (999 post burn-in for HMC, 9999 for the KSC algorithm). An argument could be made that the poor results for the KSC algorithm may be due to the chain not converging onto the target distribution. Indeed, the authors of this approach extended their Markov chain to 750,000 posterior draws (discarding the first 10,000 draws as burn-in). A fairer comparison may be to have roughly similar number of ESS for most of the estimated parameters, although this is difficult to get right with a large number of parameters to estimate. Given how well Hamiltonian Monte Carlo performed with the reparameterised model, any improvement to the off-set mixture MCMC by increasing the length of the Markov chain would make calibration just as good, but not better due to the inefficiency of the sampler. 

\subsection{Conclusion}
This research evaluated and compared different MCMC algorithms for fitting stochastic volatility models. Principled ways of checking that our algorithms are returning correct posterior estimates are important, especially as our models and algorithms become more complicated. SBC provides a general simulation design to check whether a modelling strategy is producing calibrated results for generative models.

In the context of stochastic volatility, applying Hamiltonian Monte Carlo on a reparameterised stochastic volatility model provided the most calibrated and efficient estimates when compared with KSC's algorithm applied to the approximate off-set mixture model. This also outperformed the KSC model using importance sampling weights to correct for approximation error as well as the model with non-centered parameterisation. Other parameterisations of the stochastic volatility model can be considered to see if it improves sampling performance based on these diagnostics. 

This study only considered a handful of potential algorithms, sampling strategies and parameterisations for stochastic volatility. Future research could use the same SBC design to compare and evaluate more complicated stochastic volatility models and MCMC algorithms. Overall, SBC is a valuable tool in model development and is a key part of a statistical workflow. 
 
\newpage

\bibliography{references}

\begin{thebibliography}{}

\bibitem [\protect \citeauthoryear {%
Bengtsson%
}{%
Bengtsson%
}{%
{\protect \APACyear {2021}}%
}]{%
RJ-2021-048}
\APACinsertmetastar {%
RJ-2021-048}%
\begin{APACrefauthors}%
Bengtsson, H.%
\end{APACrefauthors}%
\unskip\
\newblock
\APACrefYearMonthDay{2021}{}{}.
\newblock
{\BBOQ}\APACrefatitle {A Unifying Framework for Parallel and Distributed Processing in R using Futures} {A unifying framework for parallel and distributed processing in r using futures}.{\BBCQ}
\newblock
\APACjournalVolNumPages{The R Journal}{13}{2}{208--227}.
\newblock
\begin{APACrefURL} \url{https://doi.org/10.32614/RJ-2021-048} \end{APACrefURL}
\newblock
\begin{APACrefDOI} \doi{10.32614/RJ-2021-048} \end{APACrefDOI}
\PrintBackRefs{\CurrentBib}

\bibitem [\protect \citeauthoryear {%
Betancourt%
}{%
Betancourt%
}{%
{\protect \APACyear {2016}}%
}]{%
betancourt2016identifying}
\APACinsertmetastar {%
betancourt2016identifying}%
\begin{APACrefauthors}%
Betancourt, M.%
\end{APACrefauthors}%
\unskip\
\newblock
\APACrefYearMonthDay{2016}{}{}.
\newblock
{\BBOQ}\APACrefatitle {Identifying the optimal integration time in {H}amiltonian {M}onte {C}arlo} {Identifying the optimal integration time in {H}amiltonian {M}onte {C}arlo}.{\BBCQ}
\newblock
\APACjournalVolNumPages{arXiv preprint arXiv:1601.00225}{}{}{}.
\PrintBackRefs{\CurrentBib}

\bibitem [\protect \citeauthoryear {%
Betancourt%
}{%
Betancourt%
}{%
{\protect \APACyear {2017}}%
}]{%
betancourt2017conceptual}
\APACinsertmetastar {%
betancourt2017conceptual}%
\begin{APACrefauthors}%
Betancourt, M.%
\end{APACrefauthors}%
\unskip\
\newblock
\APACrefYearMonthDay{2017}{}{}.
\newblock
{\BBOQ}\APACrefatitle {A conceptual introduction to {H}amiltonian {M}onte {C}arlo} {A conceptual introduction to {H}amiltonian {M}onte {C}arlo}.{\BBCQ}
\newblock
\APACjournalVolNumPages{arXiv preprint arXiv:1701.02434}{}{}{}.
\PrintBackRefs{\CurrentBib}

\bibitem [\protect \citeauthoryear {%
Bürkner%
, Gabry%
, Kay%
\BCBL {}\ \BBA {} Vehtari%
}{%
Bürkner%
\ \protect \BOthers {.}}{%
{\protect \APACyear {2023}}%
}]{%
posteriorr}
\APACinsertmetastar {%
posteriorr}%
\begin{APACrefauthors}%
Bürkner, P\BHBI C.%
, Gabry, J.%
, Kay, M.%
\BCBL {}\ \BBA {} Vehtari, A.%
\end{APACrefauthors}%
\unskip\
\newblock
\APACrefYearMonthDay{2023}{}{}.
\newblock
\APACrefbtitle {posterior: Tools for Working with Posterior Distributions.} {posterior: Tools for working with posterior distributions.}
\newblock
\begin{APACrefURL} \url{https://mc-stan.org/posterior/} \end{APACrefURL}
\newblock
\APACrefnote{R package version 1.4.1}
\PrintBackRefs{\CurrentBib}

\bibitem [\protect \citeauthoryear {%
Carter%
\ \BBA {} Kohn%
}{%
Carter%
\ \BBA {} Kohn%
}{%
{\protect \APACyear {1994}}%
}]{%
carter1994gibbs}
\APACinsertmetastar {%
carter1994gibbs}%
\begin{APACrefauthors}%
Carter, C\BPBI K.%
\BCBT {}\ \BBA {} Kohn, R.%
\end{APACrefauthors}%
\unskip\
\newblock
\APACrefYearMonthDay{1994}{}{}.
\newblock
{\BBOQ}\APACrefatitle {On {G}ibbs sampling for state space models} {On {G}ibbs sampling for state space models}.{\BBCQ}
\newblock
\APACjournalVolNumPages{Biometrika}{81}{3}{541--553}.
\PrintBackRefs{\CurrentBib}

\bibitem [\protect \citeauthoryear {%
Chesney%
\ \BBA {} Scott%
}{%
Chesney%
\ \BBA {} Scott%
}{%
{\protect \APACyear {1989}}%
}]{%
chesney1989pricing}
\APACinsertmetastar {%
chesney1989pricing}%
\begin{APACrefauthors}%
Chesney, M.%
\BCBT {}\ \BBA {} Scott, L.%
\end{APACrefauthors}%
\unskip\
\newblock
\APACrefYearMonthDay{1989}{}{}.
\newblock
{\BBOQ}\APACrefatitle {Pricing {E}uropean currency options: A comparison of the modified {B}lack-{S}choles model and a random variance model} {Pricing {E}uropean currency options: A comparison of the modified {B}lack-{S}choles model and a random variance model}.{\BBCQ}
\newblock
\APACjournalVolNumPages{Journal of Financial and Quantitative Analysis}{24}{3}{267--284}.
\PrintBackRefs{\CurrentBib}

\bibitem [\protect \citeauthoryear {%
de Jong%
\ \BBA {} Shephard%
}{%
de Jong%
\ \BBA {} Shephard%
}{%
{\protect \APACyear {1995}}%
}]{%
dejong1995}
\APACinsertmetastar {%
dejong1995}%
\begin{APACrefauthors}%
de Jong, P.%
\BCBT {}\ \BBA {} Shephard, N.%
\end{APACrefauthors}%
\unskip\
\newblock
\APACrefYearMonthDay{1995}{}{}.
\newblock
{\BBOQ}\APACrefatitle {The simulation smoother for time series models} {The simulation smoother for time series models}.{\BBCQ}
\newblock
\APACjournalVolNumPages{Biometrika}{82}{}{339-350}.
\newblock
\APACrefnote{Reprinted in ``Readings in Unobserved Component Models,'' A.C. Harvey and T. Proietti, 2005, 354-367, Oxford University Press.}
\PrintBackRefs{\CurrentBib}

\bibitem [\protect \citeauthoryear {%
Duane%
, Kennedy%
, Pendleton%
\BCBL {}\ \BBA {} Roweth%
}{%
Duane%
\ \protect \BOthers {.}}{%
{\protect \APACyear {1987}}%
}]{%
duane1987hybrid}
\APACinsertmetastar {%
duane1987hybrid}%
\begin{APACrefauthors}%
Duane, S.%
, Kennedy, A\BPBI D.%
, Pendleton, B\BPBI J.%
\BCBL {}\ \BBA {} Roweth, D.%
\end{APACrefauthors}%
\unskip\
\newblock
\APACrefYearMonthDay{1987}{}{}.
\newblock
{\BBOQ}\APACrefatitle {Hybrid {M}onte {C}arlo} {Hybrid {M}onte {C}arlo}.{\BBCQ}
\newblock
\APACjournalVolNumPages{Physics Letters B}{195}{2}{216--222}.
\PrintBackRefs{\CurrentBib}

\bibitem [\protect \citeauthoryear {%
Durbin%
\ \BBA {} Koopman%
}{%
Durbin%
\ \BBA {} Koopman%
}{%
{\protect \APACyear {2012}}%
}]{%
durbin2012time}
\APACinsertmetastar {%
durbin2012time}%
\begin{APACrefauthors}%
Durbin, J.%
\BCBT {}\ \BBA {} Koopman, S\BPBI J.%
\end{APACrefauthors}%
\unskip\
\newblock
\APACrefYear{2012}.
\newblock
\APACrefbtitle {Time series analysis by state space methods} {Time series analysis by state space methods}\ (\BVOL~38).
\newblock
\APACaddressPublisher{}{OUP Oxford}.
\PrintBackRefs{\CurrentBib}

\bibitem [\protect \citeauthoryear {%
Fr{\"u}hwirth-Schnatter%
}{%
Fr{\"u}hwirth-Schnatter%
}{%
{\protect \APACyear {1995}}%
}]{%
fruhwirth1995bayesian}
\APACinsertmetastar {%
fruhwirth1995bayesian}%
\begin{APACrefauthors}%
Fr{\"u}hwirth-Schnatter, S.%
\end{APACrefauthors}%
\unskip\
\newblock
\APACrefYearMonthDay{1995}{}{}.
\newblock
{\BBOQ}\APACrefatitle {Bayesian model discrimination and {B}ayes factors for linear {G}aussian state space models} {Bayesian model discrimination and {B}ayes factors for linear {G}aussian state space models}.{\BBCQ}
\newblock
\APACjournalVolNumPages{Journal of the Royal Statistical Society: Series B (Methodological)}{57}{1}{237--246}.
\PrintBackRefs{\CurrentBib}

\bibitem [\protect \citeauthoryear {%
Fuller%
}{%
Fuller%
}{%
{\protect \APACyear {1996}}%
}]{%
fuller1996introduction}
\APACinsertmetastar {%
fuller1996introduction}%
\begin{APACrefauthors}%
Fuller, W\BPBI A.%
\end{APACrefauthors}%
\unskip\
\newblock
\APACrefYear{1996}.
\newblock
\APACrefbtitle {Introduction to time series} {Introduction to time series}\ (\PrintOrdinal{second}\ \BEd).
\newblock
\APACaddressPublisher{}{John Wiley \& Sons}.
\PrintBackRefs{\CurrentBib}

\bibitem [\protect \citeauthoryear {%
Fulton%
}{%
Fulton%
}{%
{\protect \APACyear {2018}}%
}]{%
chad2018}
\APACinsertmetastar {%
chad2018}%
\begin{APACrefauthors}%
Fulton, C.%
\end{APACrefauthors}%
\unskip\
\newblock
\APACrefYearMonthDay{2018}{}{}.
\newblock
\APACrefbtitle {Stochastic volatility: {B}ayesian inference.} {Stochastic volatility: {B}ayesian inference.}
\newblock
\APAChowpublished {\url{https://github.com/ChadFulton/tsa-notebooks/blob/master/stochastic_volatility_mcmc.ipynb}}.
\newblock
\APACaddressPublisher{}{GitHub}.
\PrintBackRefs{\CurrentBib}

\bibitem [\protect \citeauthoryear {%
Gabry%
\ \BBA {} Mahr%
}{%
Gabry%
\ \BBA {} Mahr%
}{%
{\protect \APACyear {2022}}%
}]{%
bayesplot}
\APACinsertmetastar {%
bayesplot}%
\begin{APACrefauthors}%
Gabry, J.%
\BCBT {}\ \BBA {} Mahr, T.%
\end{APACrefauthors}%
\unskip\
\newblock
\APACrefYearMonthDay{2022}{}{}.
\newblock
\APACrefbtitle {bayesplot: Plotting for Bayesian Models.} {bayesplot: Plotting for bayesian models.}
\newblock
\begin{APACrefURL} \url{https://mc-stan.org/bayesplot/} \end{APACrefURL}
\newblock
\APACrefnote{R package version 1.10.0}
\PrintBackRefs{\CurrentBib}

\bibitem [\protect \citeauthoryear {%
Gabry%
, Češnovar%
\BCBL {}\ \BBA {} Johnson%
}{%
Gabry%
\ \protect \BOthers {.}}{%
{\protect \APACyear {2022}}%
}]{%
cmdstanr}
\APACinsertmetastar {%
cmdstanr}%
\begin{APACrefauthors}%
Gabry, J.%
, Češnovar, R.%
\BCBL {}\ \BBA {} Johnson, A.%
\end{APACrefauthors}%
\unskip\
\newblock
\APACrefYearMonthDay{2022}{}{}.
\newblock
{\BBOQ}\APACrefatitle {cmdstanr: R Interface to 'CmdStan'} {cmdstanr: R interface to 'cmdstan'}{\BBCQ}\ [\bibcomputersoftwaremanual].
\newblock
\APACrefnote{https://mc-stan.org/cmdstanr/, https://discourse.mc-stan.org}
\PrintBackRefs{\CurrentBib}

\bibitem [\protect \citeauthoryear {%
Gelman%
\ \protect \BOthers {.}}{%
Gelman%
\ \protect \BOthers {.}}{%
{\protect \APACyear {2013}}%
}]{%
gelman2013bayesian}
\APACinsertmetastar {%
gelman2013bayesian}%
\begin{APACrefauthors}%
Gelman, A.%
, Carlin, J\BPBI B.%
, Stern, H\BPBI S.%
, Dunson, D\BPBI B.%
, Vehtari, A.%
\BCBL {}\ \BBA {} Rubin, D\BPBI B.%
\end{APACrefauthors}%
\unskip\
\newblock
\APACrefYear{2013}.
\newblock
\APACrefbtitle {Bayesian {D}ata {A}nalysis} {Bayesian {D}ata {A}nalysis}\ (\PrintOrdinal{third}\ \BEd).
\newblock
\APACaddressPublisher{}{Chapman \& Hall}.
\PrintBackRefs{\CurrentBib}

\bibitem [\protect \citeauthoryear {%
Gelman%
\ \BBA {} Rubin%
}{%
Gelman%
\ \BBA {} Rubin%
}{%
{\protect \APACyear {1992}}%
}]{%
gelman1992inference}
\APACinsertmetastar {%
gelman1992inference}%
\begin{APACrefauthors}%
Gelman, A.%
\BCBT {}\ \BBA {} Rubin, D\BPBI B.%
\end{APACrefauthors}%
\unskip\
\newblock
\APACrefYearMonthDay{1992}{}{}.
\newblock
{\BBOQ}\APACrefatitle {Inference from iterative simulation using multiple sequences} {Inference from iterative simulation using multiple sequences}.{\BBCQ}
\newblock
\APACjournalVolNumPages{Statistical Science}{7}{4}{457--472}.
\PrintBackRefs{\CurrentBib}

\bibitem [\protect \citeauthoryear {%
Gelman%
, Simpson%
\BCBL {}\ \BBA {} Betancourt%
}{%
Gelman%
\ \protect \BOthers {.}}{%
{\protect \APACyear {2017}}%
}]{%
gelman2017prior}
\APACinsertmetastar {%
gelman2017prior}%
\begin{APACrefauthors}%
Gelman, A.%
, Simpson, D.%
\BCBL {}\ \BBA {} Betancourt, M.%
\end{APACrefauthors}%
\unskip\
\newblock
\APACrefYearMonthDay{2017}{}{}.
\newblock
{\BBOQ}\APACrefatitle {The prior can often only be understood in the context of the likelihood} {The prior can often only be understood in the context of the likelihood}.{\BBCQ}
\newblock
\APACjournalVolNumPages{Entropy}{19}{10}{555}.
\PrintBackRefs{\CurrentBib}

\bibitem [\protect \citeauthoryear {%
Gelman%
\ \protect \BOthers {.}}{%
Gelman%
\ \protect \BOthers {.}}{%
{\protect \APACyear {2020}}%
}]{%
gelman2020bayesian}
\APACinsertmetastar {%
gelman2020bayesian}%
\begin{APACrefauthors}%
Gelman, A.%
, Vehtari, A.%
, Simpson, D.%
, Kennedy, L.%
, Margossian, C\BPBI C.%
, Carpenter, B.%
\BDBL {}Modr{\'a}k, M.%
\end{APACrefauthors}%
\unskip\
\newblock
\APACrefYearMonthDay{2020}{}{}.
\newblock
{\BBOQ}\APACrefatitle {Bayesian workflow} {Bayesian workflow}.{\BBCQ}
\newblock
\APACjournalVolNumPages{arXiv preprint arXiv:2011.01808}{}{}{}.
\PrintBackRefs{\CurrentBib}

\bibitem [\protect \citeauthoryear {%
Geman%
\ \BBA {} Geman%
}{%
Geman%
\ \BBA {} Geman%
}{%
{\protect \APACyear {1984}}%
}]{%
geman1984stochastic}
\APACinsertmetastar {%
geman1984stochastic}%
\begin{APACrefauthors}%
Geman, S.%
\BCBT {}\ \BBA {} Geman, D.%
\end{APACrefauthors}%
\unskip\
\newblock
\APACrefYearMonthDay{1984}{}{}.
\newblock
{\BBOQ}\APACrefatitle {Stochastic relaxation, {G}ibbs distributions, and the {B}ayesian restoration of images} {Stochastic relaxation, {G}ibbs distributions, and the {B}ayesian restoration of images}.{\BBCQ}
\newblock
\APACjournalVolNumPages{IEEE Transactions on pattern analysis and machine intelligence}{}{6}{721--741}.
\PrintBackRefs{\CurrentBib}

\bibitem [\protect \citeauthoryear {%
Geyer%
}{%
Geyer%
}{%
{\protect \APACyear {1992}}%
}]{%
geyer1992practical}
\APACinsertmetastar {%
geyer1992practical}%
\begin{APACrefauthors}%
Geyer, C\BPBI J.%
\end{APACrefauthors}%
\unskip\
\newblock
\APACrefYearMonthDay{1992}{}{}.
\newblock
{\BBOQ}\APACrefatitle {Practical {M}arkov chain {M}onte {C}arlo} {Practical {M}arkov chain {M}onte {C}arlo}.{\BBCQ}
\newblock
\APACjournalVolNumPages{Statistical Science}{}{}{473--483}.
\PrintBackRefs{\CurrentBib}

\bibitem [\protect \citeauthoryear {%
Harris%
\ \protect \BOthers {.}}{%
Harris%
\ \protect \BOthers {.}}{%
{\protect \APACyear {2020}}%
}]{%
harris2020array}
\APACinsertmetastar {%
harris2020array}%
\begin{APACrefauthors}%
Harris, C\BPBI R.%
, Millman, K\BPBI J.%
, van~der Walt, S\BPBI J.%
, Gommers, R.%
, Virtanen, P.%
, Cournapeau, D.%
\BDBL {}Oliphant, T\BPBI E.%
\end{APACrefauthors}%
\unskip\
\newblock
\APACrefYearMonthDay{2020}{{\APACmonth{09}}}{}.
\newblock
{\BBOQ}\APACrefatitle {Array programming with {NumPy}} {Array programming with {NumPy}}.{\BBCQ}
\newblock
\APACjournalVolNumPages{Nature}{585}{7825}{357--362}.
\newblock
\begin{APACrefURL} \url{https://doi.org/10.1038/s41586-020-2649-2} \end{APACrefURL}
\newblock
\begin{APACrefDOI} \doi{10.1038/s41586-020-2649-2} \end{APACrefDOI}
\PrintBackRefs{\CurrentBib}

\bibitem [\protect \citeauthoryear {%
Hastings%
}{%
Hastings%
}{%
{\protect \APACyear {1970}}%
}]{%
hastings1970monte}
\APACinsertmetastar {%
hastings1970monte}%
\begin{APACrefauthors}%
Hastings, W\BPBI K.%
\end{APACrefauthors}%
\unskip\
\newblock
\APACrefYearMonthDay{1970}{}{}.
\newblock
{\BBOQ}\APACrefatitle {Monte {C}arlo sampling methods using {M}arkov chains and their applications} {Monte {C}arlo sampling methods using {M}arkov chains and their applications}.{\BBCQ}
\newblock

\PrintBackRefs{\CurrentBib}

\bibitem [\protect \citeauthoryear {%
Hoffman%
\ \BBA {} Gelman%
}{%
Hoffman%
\ \BBA {} Gelman%
}{%
{\protect \APACyear {2014}}%
}]{%
hoffman2014no}
\APACinsertmetastar {%
hoffman2014no}%
\begin{APACrefauthors}%
Hoffman, M\BPBI D.%
\BCBT {}\ \BBA {} Gelman, A.%
\end{APACrefauthors}%
\unskip\
\newblock
\APACrefYearMonthDay{2014}{}{}.
\newblock
{\BBOQ}\APACrefatitle {The {N}o-{U}-{T}urn {S}ampler: adaptively setting path lengths in {H}amiltonian {M}onte {C}arlo.} {The {N}o-{U}-{T}urn {S}ampler: adaptively setting path lengths in {H}amiltonian {M}onte {C}arlo.}{\BBCQ}
\newblock
\APACjournalVolNumPages{Journal of Machine Learing Research}{15}{1}{1593--1623}.
\PrintBackRefs{\CurrentBib}

\bibitem [\protect \citeauthoryear {%
Hull%
\ \BBA {} White%
}{%
Hull%
\ \BBA {} White%
}{%
{\protect \APACyear {1987}}%
}]{%
hull1987pricing}
\APACinsertmetastar {%
hull1987pricing}%
\begin{APACrefauthors}%
Hull, J.%
\BCBT {}\ \BBA {} White, A.%
\end{APACrefauthors}%
\unskip\
\newblock
\APACrefYearMonthDay{1987}{}{}.
\newblock
{\BBOQ}\APACrefatitle {The pricing of options on assets with stochastic volatilities} {The pricing of options on assets with stochastic volatilities}.{\BBCQ}
\newblock
\APACjournalVolNumPages{The Journal of Finance}{42}{2}{281--300}.
\PrintBackRefs{\CurrentBib}

\bibitem [\protect \citeauthoryear {%
Kalman%
}{%
Kalman%
}{%
{\protect \APACyear {1960}}%
}]{%
kalman1960new}
\APACinsertmetastar {%
kalman1960new}%
\begin{APACrefauthors}%
Kalman, R\BPBI E.%
\end{APACrefauthors}%
\unskip\
\newblock
\APACrefYearMonthDay{1960}{}{}.
\newblock
{\BBOQ}\APACrefatitle {A new approach to linear filtering and prediction problems} {A new approach to linear filtering and prediction problems}.{\BBCQ}
\newblock
\APACjournalVolNumPages{Journal of Basic Engineering}{82}{}{35--45}.
\PrintBackRefs{\CurrentBib}

\bibitem [\protect \citeauthoryear {%
Kim%
, Shephard%
\BCBL {}\ \BBA {} Chib%
}{%
Kim%
\ \protect \BOthers {.}}{%
{\protect \APACyear {1998}}%
}]{%
kim1998stochastic}
\APACinsertmetastar {%
kim1998stochastic}%
\begin{APACrefauthors}%
Kim, S.%
, Shephard, N.%
\BCBL {}\ \BBA {} Chib, S.%
\end{APACrefauthors}%
\unskip\
\newblock
\APACrefYearMonthDay{1998}{}{}.
\newblock
{\BBOQ}\APACrefatitle {Stochastic volatility: likelihood inference and comparison with {ARCH} models} {Stochastic volatility: likelihood inference and comparison with {ARCH} models}.{\BBCQ}
\newblock
\APACjournalVolNumPages{The Review of Economic Studies}{65}{3}{361--393}.
\PrintBackRefs{\CurrentBib}

\bibitem [\protect \citeauthoryear {%
Koopman%
, Shephard%
\BCBL {}\ \BBA {} Doornik%
}{%
Koopman%
\ \protect \BOthers {.}}{%
{\protect \APACyear {1996}}%
}]{%
koopman1996ssfpack}
\APACinsertmetastar {%
koopman1996ssfpack}%
\begin{APACrefauthors}%
Koopman, S\BPBI J.%
, Shephard, N.%
\BCBL {}\ \BBA {} Doornik, J\BPBI A.%
\end{APACrefauthors}%
\unskip\
\newblock
\APACrefYearMonthDay{1996}{}{}.
\newblock
\APACrefbtitle {SsfPack 1.1: filtering, smoothing and simulation algorithms for state space models in {O}x.} {Ssfpack 1.1: filtering, smoothing and simulation algorithms for state space models in {O}x.}
\PrintBackRefs{\CurrentBib}

\bibitem [\protect \citeauthoryear {%
Metropolis%
, Rosenbluth%
, Rosenbluth%
, Teller%
\BCBL {}\ \BBA {} Teller%
}{%
Metropolis%
\ \protect \BOthers {.}}{%
{\protect \APACyear {1953}}%
}]{%
metropolis1953equation}
\APACinsertmetastar {%
metropolis1953equation}%
\begin{APACrefauthors}%
Metropolis, N.%
, Rosenbluth, A\BPBI W.%
, Rosenbluth, M\BPBI N.%
, Teller, A\BPBI H.%
\BCBL {}\ \BBA {} Teller, E.%
\end{APACrefauthors}%
\unskip\
\newblock
\APACrefYearMonthDay{1953}{}{}.
\newblock
{\BBOQ}\APACrefatitle {Equation of state calculations by fast computing machines} {Equation of state calculations by fast computing machines}.{\BBCQ}
\newblock
\APACjournalVolNumPages{The journal of chemical physics}{21}{6}{1087--1092}.
\PrintBackRefs{\CurrentBib}

\bibitem [\protect \citeauthoryear {%
Neal%
}{%
Neal%
}{%
{\protect \APACyear {1995}}%
}]{%
neal1995bayesian}
\APACinsertmetastar {%
neal1995bayesian}%
\begin{APACrefauthors}%
Neal, R\BPBI M.%
\end{APACrefauthors}%
\unskip\
\newblock
\APACrefYear{1995}.
\unskip\
\newblock
\APACrefbtitle {{B}ayesian {L}earning for {N}eural {N}etworks} {{B}ayesian {L}earning for {N}eural {N}etworks}\ \APACtypeAddressSchool {\BUPhD}{}{}.
\unskip\
\newblock
\APACaddressSchool {}{University of Toronto}.
\PrintBackRefs{\CurrentBib}

\bibitem [\protect \citeauthoryear {%
Neal%
}{%
Neal%
}{%
{\protect \APACyear {2003}}%
}]{%
neal2003slice}
\APACinsertmetastar {%
neal2003slice}%
\begin{APACrefauthors}%
Neal, R\BPBI M.%
\end{APACrefauthors}%
\unskip\
\newblock
\APACrefYearMonthDay{2003}{}{}.
\newblock
{\BBOQ}\APACrefatitle {Slice sampling} {Slice sampling}.{\BBCQ}
\newblock
\APACjournalVolNumPages{The Annals of Statistics}{31}{3}{705--767}.
\PrintBackRefs{\CurrentBib}

\bibitem [\protect \citeauthoryear {%
Neal%
}{%
Neal%
}{%
{\protect \APACyear {2011}}%
}]{%
neal2011mcmc}
\APACinsertmetastar {%
neal2011mcmc}%
\begin{APACrefauthors}%
Neal, R\BPBI M.%
\end{APACrefauthors}%
\unskip\
\newblock
\APACrefYearMonthDay{2011}{}{}.
\newblock
{\BBOQ}\APACrefatitle {{MCMC} using {H}amiltonian dynamics} {{MCMC} using {H}amiltonian dynamics}.{\BBCQ}
\newblock
\APACjournalVolNumPages{Handbook of Markov Chain Monte Carlo}{2}{11}{2}.
\PrintBackRefs{\CurrentBib}

\bibitem [\protect \citeauthoryear {%
Ooms%
}{%
Ooms%
}{%
{\protect \APACyear {2014}}%
}]{%
jsonlite}
\APACinsertmetastar {%
jsonlite}%
\begin{APACrefauthors}%
Ooms, J.%
\end{APACrefauthors}%
\unskip\
\newblock
\APACrefYearMonthDay{2014}{}{}.
\newblock
{\BBOQ}\APACrefatitle {The jsonlite Package: A Practical and Consistent Mapping Between JSON Data and R Objects} {The jsonlite package: A practical and consistent mapping between json data and r objects}.{\BBCQ}
\newblock
\APACjournalVolNumPages{arXiv:1403.2805 [stat.CO]}{}{}{}.
\newblock
\begin{APACrefURL} \url{https://arxiv.org/abs/1403.2805} \end{APACrefURL}
\PrintBackRefs{\CurrentBib}

\bibitem [\protect \citeauthoryear {%
Oriol%
\ \protect \BOthers {.}}{%
Oriol%
\ \protect \BOthers {.}}{%
{\protect \APACyear {2023}}%
}]{%
pymc2023}
\APACinsertmetastar {%
pymc2023}%
\begin{APACrefauthors}%
Oriol, A\BHBI P.%
, Virgile, A.%
, Colin, C.%
, Larry, D.%
, J., F\BPBI C.%
, Maxim, K.%
\BDBL {}Robert, Z.%
\end{APACrefauthors}%
\unskip\
\newblock
\APACrefYearMonthDay{2023}{}{}.
\newblock
{\BBOQ}\APACrefatitle {{PyMC}: A Modern and Comprehensive Probabilistic Programming Framework in {P}ython} {{PyMC}: A modern and comprehensive probabilistic programming framework in {P}ython}.{\BBCQ}
\newblock
\APACjournalVolNumPages{{Peer J} Computer Science}{9}{}{e1516}.
\newblock
\begin{APACrefDOI} \doi{10.7717/peerj-cs.1516} \end{APACrefDOI}
\PrintBackRefs{\CurrentBib}

\bibitem [\protect \citeauthoryear {%
{R Core Team}%
}{%
{R Core Team}%
}{%
{\protect \APACyear {2023}}%
}]{%
rlang}
\APACinsertmetastar {%
rlang}%
\begin{APACrefauthors}%
{R Core Team}.%
\end{APACrefauthors}%
\unskip\
\newblock
\APACrefYearMonthDay{2023}{}{}.
\newblock
{\BBOQ}\APACrefatitle {R: A Language and Environment for Statistical Computing} {R: A language and environment for statistical computing}{\BBCQ}\ [\bibcomputersoftwaremanual].
\newblock
\APACaddressPublisher{Vienna, Austria}{}.
\newblock
\begin{APACrefURL} \url{https://www.R-project.org/} \end{APACrefURL}
\PrintBackRefs{\CurrentBib}

\bibitem [\protect \citeauthoryear {%
Richardson%
\ \protect \BOthers {.}}{%
Richardson%
\ \protect \BOthers {.}}{%
{\protect \APACyear {2023}}%
}]{%
arrow}
\APACinsertmetastar {%
arrow}%
\begin{APACrefauthors}%
Richardson, N.%
, Cook, I.%
, Crane, N.%
, Dunnington, D.%
, François, R.%
, Keane, J.%
\BDBL {}{Apache Arrow}%
\end{APACrefauthors}%
\unskip\
\newblock
\APACrefYearMonthDay{2023}{}{}.
\newblock
{\BBOQ}\APACrefatitle {arrow: Integration to 'Apache' 'Arrow'} {arrow: Integration to 'apache' 'arrow'}{\BBCQ}\ [\bibcomputersoftwaremanual].
\newblock
\APACrefnote{https://github.com/apache/arrow/, https://arrow.apache.org/docs/r/}
\PrintBackRefs{\CurrentBib}

\bibitem [\protect \citeauthoryear {%
Ryan%
\ \BBA {} Ulrich%
}{%
Ryan%
\ \BBA {} Ulrich%
}{%
{\protect \APACyear {2022}}%
}]{%
quantmod}
\APACinsertmetastar {%
quantmod}%
\begin{APACrefauthors}%
Ryan, J\BPBI A.%
\BCBT {}\ \BBA {} Ulrich, J\BPBI M.%
\end{APACrefauthors}%
\unskip\
\newblock
\APACrefYearMonthDay{2022}{}{}.
\newblock
{\BBOQ}\APACrefatitle {quantmod: Quantitative Financial Modelling Framework} {quantmod: Quantitative financial modelling framework}{\BBCQ}\ [\bibcomputersoftwaremanual].
\newblock
\begin{APACrefURL} \url{https://CRAN.R-project.org/package=quantmod} \end{APACrefURL}
\newblock
\APACrefnote{R package version 0.4.20}
\PrintBackRefs{\CurrentBib}

\bibitem [\protect \citeauthoryear {%
Seabold%
\ \BBA {} Perktold%
}{%
Seabold%
\ \BBA {} Perktold%
}{%
{\protect \APACyear {2010}}%
}]{%
seabold2010statsmodels}
\APACinsertmetastar {%
seabold2010statsmodels}%
\begin{APACrefauthors}%
Seabold, S.%
\BCBT {}\ \BBA {} Perktold, J.%
\end{APACrefauthors}%
\unskip\
\newblock
\APACrefYearMonthDay{2010}{}{}.
\newblock
{\BBOQ}\APACrefatitle {Statsmodels: {E}conometric and statistical modeling with python} {Statsmodels: {E}conometric and statistical modeling with python}.{\BBCQ}
\newblock
\BIn{} \APACrefbtitle {9th Python in Science Conference.} {9th python in science conference.}
\PrintBackRefs{\CurrentBib}

\bibitem [\protect \citeauthoryear {%
{Stan Development Team}%
}{%
{Stan Development Team}%
}{%
{\protect \APACyear {2023}}%
}]{%
stan}
\APACinsertmetastar {%
stan}%
\begin{APACrefauthors}%
{Stan Development Team}.%
\end{APACrefauthors}%
\unskip\
\newblock
\APACrefYearMonthDay{2023}{}{}.
\newblock
\APACrefbtitle {Stan {M}odeling {L}anguage {U}sers {G}uide and {R}eference {M}anual.} {Stan {M}odeling {L}anguage {U}sers {G}uide and {R}eference {M}anual.}
\newblock
\begin{APACrefURL} \url{https://mc-stan.org/users/documentation/} \end{APACrefURL}
\PrintBackRefs{\CurrentBib}

\bibitem [\protect \citeauthoryear {%
Strickland%
, Martin%
\BCBL {}\ \BBA {} Forbes%
}{%
Strickland%
\ \protect \BOthers {.}}{%
{\protect \APACyear {2008}}%
}]{%
strickland2008parameterisation}
\APACinsertmetastar {%
strickland2008parameterisation}%
\begin{APACrefauthors}%
Strickland, C\BPBI M.%
, Martin, G\BPBI M.%
\BCBL {}\ \BBA {} Forbes, C\BPBI S.%
\end{APACrefauthors}%
\unskip\
\newblock
\APACrefYearMonthDay{2008}{}{}.
\newblock
{\BBOQ}\APACrefatitle {Parameterisation and efficient {MCMC} estimation of non-{G}aussian state space models} {Parameterisation and efficient {MCMC} estimation of non-{G}aussian state space models}.{\BBCQ}
\newblock
\APACjournalVolNumPages{Computational Statistics \& Data Analysis}{52}{6}{2911--2930}.
\PrintBackRefs{\CurrentBib}

\bibitem [\protect \citeauthoryear {%
Talts%
, Betancourt%
, Simpson%
, Vehtari%
\BCBL {}\ \BBA {} Gelman%
}{%
Talts%
\ \protect \BOthers {.}}{%
{\protect \APACyear {2020}}%
}]{%
talts2020validating}
\APACinsertmetastar {%
talts2020validating}%
\begin{APACrefauthors}%
Talts, S.%
, Betancourt, M.%
, Simpson, D.%
, Vehtari, A.%
\BCBL {}\ \BBA {} Gelman, A.%
\end{APACrefauthors}%
\unskip\
\newblock
\APACrefYearMonthDay{2020}{}{}.
\newblock
{\BBOQ}\APACrefatitle {Validating {B}ayesian inference algorithms with simulation-based calibration} {Validating {B}ayesian inference algorithms with simulation-based calibration}.{\BBCQ}
\newblock
\APACjournalVolNumPages{arXiv preprint arXiv:1804.06788}{}{}{}.
\PrintBackRefs{\CurrentBib}

\bibitem [\protect \citeauthoryear {%
Van~Rossum%
\ \BBA {} Drake%
}{%
Van~Rossum%
\ \BBA {} Drake%
}{%
{\protect \APACyear {2009}}%
}]{%
10.5555/1593511}
\APACinsertmetastar {%
10.5555/1593511}%
\begin{APACrefauthors}%
Van~Rossum, G.%
\BCBT {}\ \BBA {} Drake, F\BPBI L.%
\end{APACrefauthors}%
\unskip\
\newblock
\APACrefYear{2009}.
\newblock
\APACrefbtitle {Python 3 Reference Manual} {Python 3 reference manual}.
\newblock
\APACaddressPublisher{Scotts Valley, CA}{CreateSpace}.
\PrintBackRefs{\CurrentBib}

\bibitem [\protect \citeauthoryear {%
Vehtari%
, Gelman%
, Simpson%
, Carpenter%
\BCBL {}\ \BBA {} B{\"u}rkner%
}{%
Vehtari%
\ \protect \BOthers {.}}{%
{\protect \APACyear {2021}}%
}]{%
vehtari2021rank}
\APACinsertmetastar {%
vehtari2021rank}%
\begin{APACrefauthors}%
Vehtari, A.%
, Gelman, A.%
, Simpson, D.%
, Carpenter, B.%
\BCBL {}\ \BBA {} B{\"u}rkner, P\BHBI C.%
\end{APACrefauthors}%
\unskip\
\newblock
\APACrefYearMonthDay{2021}{}{}.
\newblock
{\BBOQ}\APACrefatitle {Rank-normalization, folding, and localization: An improved $\widehat{R}$ for assessing convergence of {MCMC} (with discussion)} {Rank-normalization, folding, and localization: An improved $\widehat{R}$ for assessing convergence of {MCMC} (with discussion)}.{\BBCQ}
\newblock
\APACjournalVolNumPages{Bayesian Analysis}{16}{2}{667--718}.
\PrintBackRefs{\CurrentBib}

\bibitem [\protect \citeauthoryear {%
Virtanen%
\ \protect \BOthers {.}}{%
Virtanen%
\ \protect \BOthers {.}}{%
{\protect \APACyear {2020}}%
}]{%
2020SciPy-NMeth}
\APACinsertmetastar {%
2020SciPy-NMeth}%
\begin{APACrefauthors}%
Virtanen, P.%
, Gommers, R.%
, Oliphant, T\BPBI E.%
, Haberland, M.%
, Reddy, T.%
, Cournapeau, D.%
\BDBL {}{SciPy 1.0 Contributors}%
\end{APACrefauthors}%
\unskip\
\newblock
\APACrefYearMonthDay{2020}{}{}.
\newblock
{\BBOQ}\APACrefatitle {{{SciPy} 1.0: Fundamental Algorithms for Scientific Computing in Python}} {{{SciPy} 1.0: Fundamental Algorithms for Scientific Computing in Python}}.{\BBCQ}
\newblock
\APACjournalVolNumPages{Nature Methods}{17}{}{261--272}.
\newblock
\begin{APACrefDOI} \doi{10.1038/s41592-019-0686-2} \end{APACrefDOI}
\PrintBackRefs{\CurrentBib}

\bibitem [\protect \citeauthoryear {%
{W}es {M}c{K}inney%
}{%
{W}es {M}c{K}inney%
}{%
{\protect \APACyear {2010}}%
}]{%
mckinney-proc-scipy-2010}
\APACinsertmetastar {%
mckinney-proc-scipy-2010}%
\begin{APACrefauthors}%
{W}es {M}c{K}inney.%
\end{APACrefauthors}%
\unskip\
\newblock
\APACrefYearMonthDay{2010}{}{}.
\newblock
{\BBOQ}\APACrefatitle {{D}ata {S}tructures for {S}tatistical {C}omputing in {P}ython} {{D}ata {S}tructures for {S}tatistical {C}omputing in {P}ython}.{\BBCQ}
\newblock
\BIn{} {S}t\'efan van~der {W}alt\ \BBA {} {J}arrod {M}illman\ (\BEDS), \APACrefbtitle {{P}roceedings of the 9th {P}ython in {S}cience {C}onference} {{P}roceedings of the 9th {P}ython in {S}cience {C}onference}\ (\BPG~56 - 61).
\newblock
\begin{APACrefDOI} \doi{10.25080/Majora-92bf1922-00a} \end{APACrefDOI}
\PrintBackRefs{\CurrentBib}

\bibitem [\protect \citeauthoryear {%
Wickham%
\ \protect \BOthers {.}}{%
Wickham%
\ \protect \BOthers {.}}{%
{\protect \APACyear {2019}}%
}]{%
tidyverse}
\APACinsertmetastar {%
tidyverse}%
\begin{APACrefauthors}%
Wickham, H.%
, Averick, M.%
, Bryan, J.%
, Chang, W.%
, McGowan, L\BPBI D.%
, François, R.%
\BDBL {}Yutani, H.%
\end{APACrefauthors}%
\unskip\
\newblock
\APACrefYearMonthDay{2019}{}{}.
\newblock
{\BBOQ}\APACrefatitle {Welcome to the {tidyverse}} {Welcome to the {tidyverse}}.{\BBCQ}
\newblock
\APACjournalVolNumPages{Journal of Open Source Software}{4}{43}{1686}.
\newblock
\begin{APACrefDOI} \doi{10.21105/joss.01686} \end{APACrefDOI}
\PrintBackRefs{\CurrentBib}

\end{thebibliography}

\newpage

\section{Appendix A: Mixture Gaussian weights}

\begin{table}[H]
    \centering
    \begin{tabular}{lccc} 
          $\omega$ &$Pr(\omega = i)$&  $m_i$&  $\nu^2_i$\\ 
          1&0.00730  &  -10.12999&  5.79596\\ 
          2&0.10556  &   -3.97281 &  2.61369\\ 
          3&0.00002 &  -8.56686 &   5.17950\\ 
          4&0.04395 &  2.77786  &   0.16735 \\ 
          5&0.34001&   0.61942    &  0.64009\\ 
          6&0.24566 &  1.79518    &  0.34023 \\ 
          7&0.25750 &  -1.08819    &  1.26261\\ 
    \end{tabular} 
\end{table}

\newpage

\section{Appendix B: Sampling from Gaussian off-set mixture model}
    The steps in the MCMC process are described in Algorithm \ref{alg:ksc}. The initial values are set to the values outlined in \citet{kim1998stochastic}, except for the mixing indicator which is unspecified. This is arbitrarily set at the 4th Gaussian density found in Appendix A. 
        
    \subsection*{Sampling mixture density}
    For sampling the mixture model the Gaussian mixture density is rewritten with respect to a indicator variable $s_t$
        \begin{align}
        &z_t | s_t = i \sim N(m_i - 1.2704, \nu^2) \\
        &Pr(s_t = i) = q_i.
        \end{align}
        The vector $s$ indexes the seven mixture of Gaussians and a component indicator variable for each t, $s_t$, is sampled from probability mass function: 
        \begin{align}
        Pr(s_t = i | y_t^{\ast}, h_t) \propto q_i f_N(y_t^{\ast} | h_t + m_t - 1.2704, \nu^2).
        \end{align}

\subsection*{Conjugate posterior distributions}

\textbf{Sampling} $\boldsymbol{\sigma_{\eta}^2}$

Inverse gamma conjugate posterior distribution:

$$
\sigma^2_{\eta} | y,h,\phi,\mu \sim IG \Bigl\{\frac{n+\sigma_r}{2}, \frac{0.05 + (h_1 - \mu)^2 (1 - \phi^2) + \sum_{t=1}^{n-1}((h_{t+1} - \mu) - \phi(h_t - \mu))^2}{2}\Bigr\}
$$

\textbf{Sampling}  $\boldsymbol{\mu}$

Gaussian conjugate posterior distribution:

$$
\mu | h,\phi,\sigma^2_{\eta}  \sim N(\hat{\mu}, \sigma^2_{\mu})
$$

Where

$$
\begin{aligned}
\hat{\mu} &= \sigma^2_{\mu} \Bigl\{\frac{(1-\phi^2)}{\sigma_{\eta}^2}h_1 +\frac{(1-\phi^2)}{\sigma_{\eta}^2} \sum_{t=1}^{n-1} (h_{t+1} - \phi h_t)\Bigr\} \\
\sigma^2_{\mu} &= \sigma^2_{\eta} \{(n-1)(1-\phi)^2 + (1-\phi^2)\}^{-1}
\end{aligned}
$$

\subsection*{Metropolis Hastings Step}
\textbf{Sampling}  $\boldsymbol{\phi}$

Metropolis Hastings accept/reject procedure:

1) Generate proposal $\phi^\ast$ from $N(\hat{\phi}, V_{\phi})$ where $\hat{\phi} = \frac{\sum_{t=1}^{n-1} (h_{t+1} - \mu)(h_t - \mu)}{\sum_{t=1}^{n-1} (h_t - \mu)^2}$ and $V_{\phi} = \sigma^2_{\eta} \{\sum_{t=1}^{n-1} (h_t - \mu)^2\}^{-1}$

2) Accept proposal as $\phi^{(i)}$ with probability $e^{\{g(\phi^\ast) - g(\phi^{(i-1)}\}}$ such that $g(\phi) = \log (\pi (\phi)) - \frac {(h_t - \mu)^2 (1-\phi^2)}{2 \sigma_{\eta}^2} + \frac{1}{2} \log (1-\phi^2)$

        \begin{algorithm}[H]
            \caption{KSC MCMC Algorithm}\label{alg:ksc}
            \begin{algorithmic}
            \Require $s_0 = 4$, $\mu_0 = 0$, $\phi_0 = 0.95$, $\sigma^{2}_{\eta,0} = 0.02$
            \For{\texttt{b in} $1:B_{draws}$}
                    \State \text{Sample states (Kalman Filter and Smoother): } $\boldsymbol{h}_b \sim h|y^{\ast}, s_{b-1}, \phi_{b-1}, \sigma^{2}_{\eta,b-1}, \mu_{b-1}$ 
                    \State \text{Sample mixture indicators: } $s_b \sim s|y^{\ast}, \boldsymbol{h}_{b-1}$
                    \State \text{Sample from conjugate density $\mu$: } $\mu_b \sim \mu|y_{\ast}, s_{b-1}, \phi_{b-1}, \sigma^{2}_{\eta, b-1}, \boldsymbol{h}_{b-1}$
                    \State \text{Sample from conjugate density $\sigma^2_{\eta}$: } $\mu_b \sim \mu|y^{\ast}, s_{b-1}, \phi_{b-1}, \mu_{b-1}, \boldsymbol{h}_{b-1}$
                    \State \text{Sample via Metropolis-Hastings $\phi$: } $\phi_b \sim \phi|y^{\ast}, s_{b-1}, \mu_{b-1}, \sigma^{2}_{\eta, b-1}, \boldsymbol{h}_{b-1}$
                  \EndFor
            \end{algorithmic}
            \end{algorithm}

\newpage

\section{Appendix C: Hamiltonian Monte Carlo algorithm description}
Let $\theta$ be a target parameter and $\alpha$ be the auxiliary momentum variable. The Hamiltonian is defined as
\begin{align}
H(\theta, \alpha) \equiv - \log \pi(\theta, \alpha).
\end{align}

This can be broken down into kinetic energy $K(\theta, \alpha)$, the density over the auxiliary momentum, and potential energy $V(\theta)$, the density of the target posterior distribution
\begin{align}
H(\theta, \alpha) &= - \log \pi(\alpha | \theta) - \log \pi(\theta) \\ 
&\equiv  K(\theta, \alpha) + V(\theta).
\end{align}

Starting with an initial draw of $\theta$ (user defined or randomly generated), the HMC iteration or proposal is determined by 3 steps.

\textbf{1) Randomly sample a momentum value}
\begin{align}
\alpha\sim \mathrm{Multinormal}(0, M),
\end{align}

where M is some diagonal mass matrix (assuming independence between momentum variables).

\textbf{2) Solve the set of Hamiltonian (differential) equations which generates a proposal} $(\theta^{\ast}, \alpha^{\ast})$ given by
\begin{align}
\frac{\partial \theta}{\partial t} &= \frac{\partial H}{\partial \alpha} = \frac{\partial K}{\partial \alpha} \\
\frac{\partial \alpha}{\partial t} &= \frac{\partial H}{\partial \theta} = - \frac{\partial K}{\partial \theta} - \frac{\partial V}        {\partial \theta},
\end{align}

where $\frac{\partial V}{\partial \theta}$ is the gradient of the target posterior density. The solution to these differential equations is approximated using a Leapfrog integrator which gives a discrete approximate solution. The parameters which are tuned in this process are the number of steps $G$ and the size of the steps $\epsilon$. The Leapfrog algorithm performs a half step of $\alpha$ and then a full step of the parameter $\theta$, and another half step of the momentum variable $\alpha$ and repeats this $G$ times where
\begin{align}
\alpha &\leftarrow \alpha - \frac{\epsilon}{2} \frac{\partial V}{\partial \theta} \\
\theta &\leftarrow \theta + \epsilon M^{-1} \alpha \\
\alpha &\leftarrow \alpha - \frac{\epsilon}{2} \frac{\partial V}{\partial \theta}.
\end{align}

The final step is the proposal $(\theta^{\ast}, \alpha^{\ast})$.

\textbf{3) Metropolis Hastings Accept/Reject}

Let $(\theta^{b-1}, \alpha^{b-1})$ be the values before the Leapfrog integrator, then
\begin{align}
a = \frac{\pi(\theta^{\ast} | y) \pi(\alpha^{\ast})}{\pi(\theta^{b-1} | y) \pi(\alpha^{b-1})},
\end{align}

with sampled value
\begin{align}
\theta^b = \begin{cases}
    \theta^{\ast},& \text{with probability } min(a,1)\\
    \theta^{b-1}, & \text{otherwise}.
    \end{cases}
\end{align}

\section{Appendix D: 1000 SBC iteration results}

    \begin{figure}[H]
        \centering
        \includegraphics[scale=0.09]{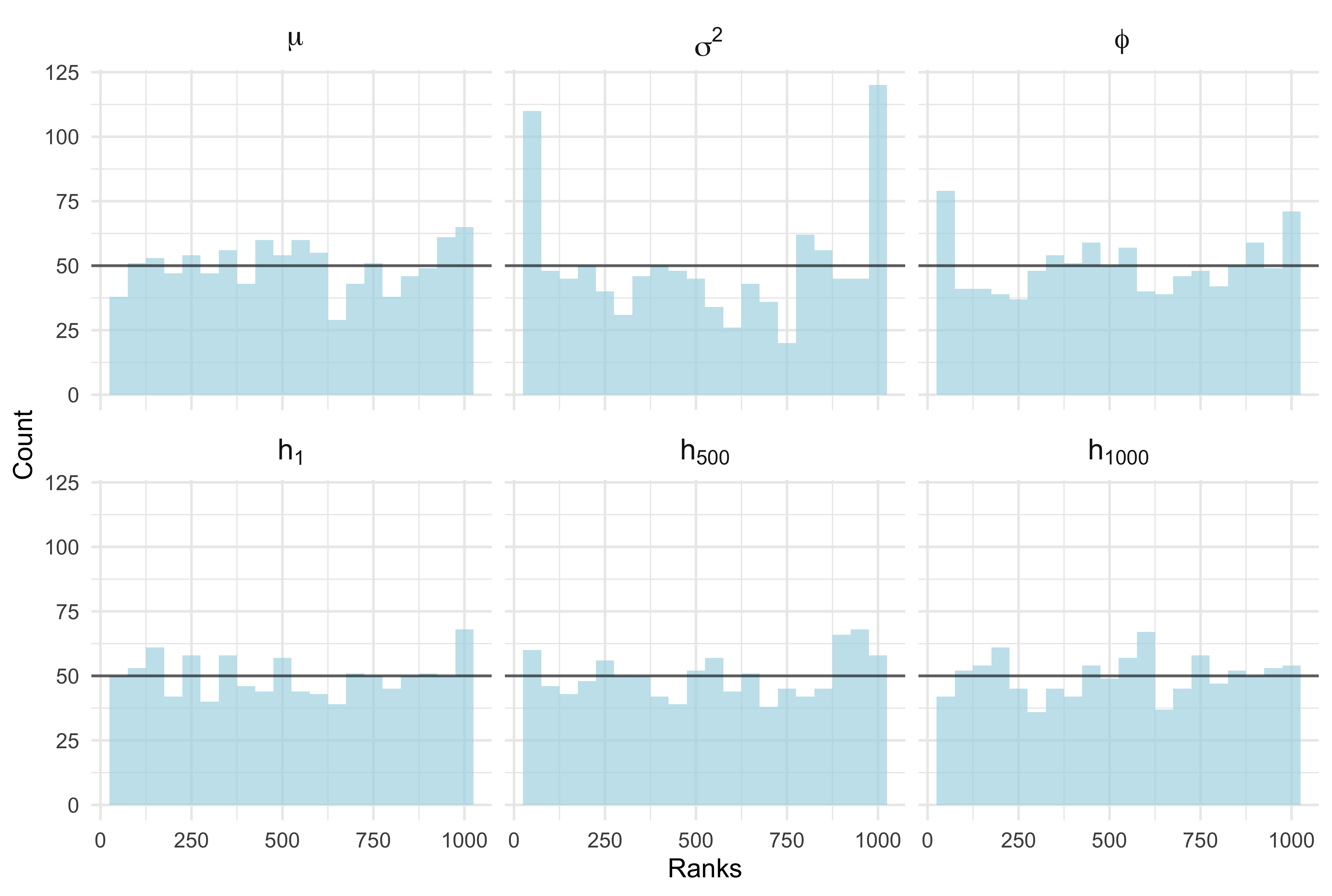}
        \caption{\textbf{1000 SBC iterations for centered model using Hamiltonian Monte Carlo}. $\sigma^2$ displays inflated frequencies on both ends of the histogram and $\phi$ shows inflated frequency left hand side. This suggests that the average posterior samples of $\sigma^2$ are under-dispersed relative to the prior distribution and posterior samples of $\phi$ are overestimating the true parameter.}
        \label{fig:cphmc1k}
    \end{figure} 

        \begin{figure}[H]
        \centering
        \includegraphics[scale=0.09]{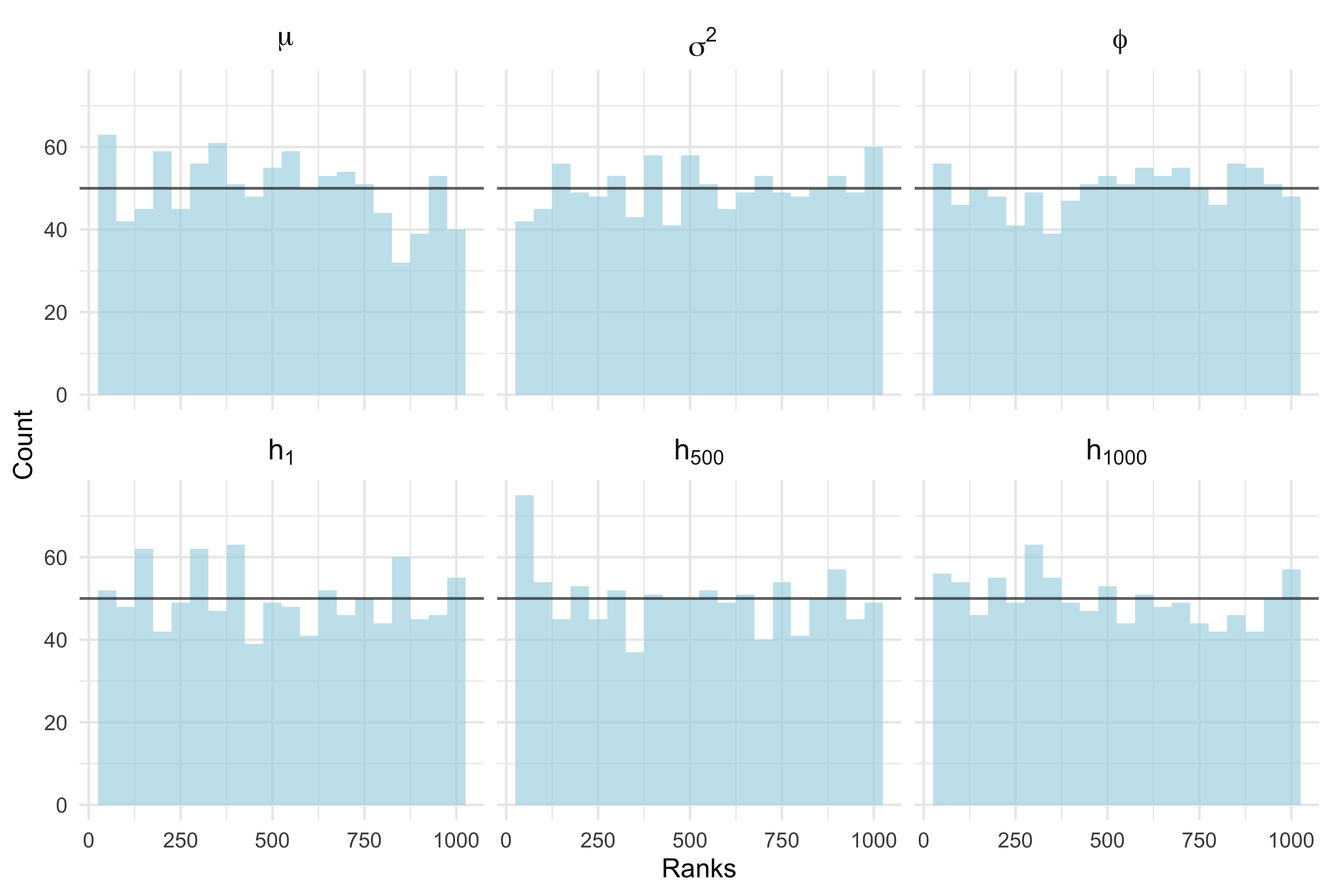}
        \caption{\textbf{1000 SBC iterations for reparameterised model using Hamiltonian Monte Carlo}. The rank statistic distributions for $\sigma^2$ and $\phi$ have improved. There is some evidence that the posterior of the 500th state variable is overestimating the distribution of its true value.}
        \label{fig:ncphmc1k}
    \end{figure}

        \begin{figure}[H]
        \centering
        \includegraphics[scale=0.09]{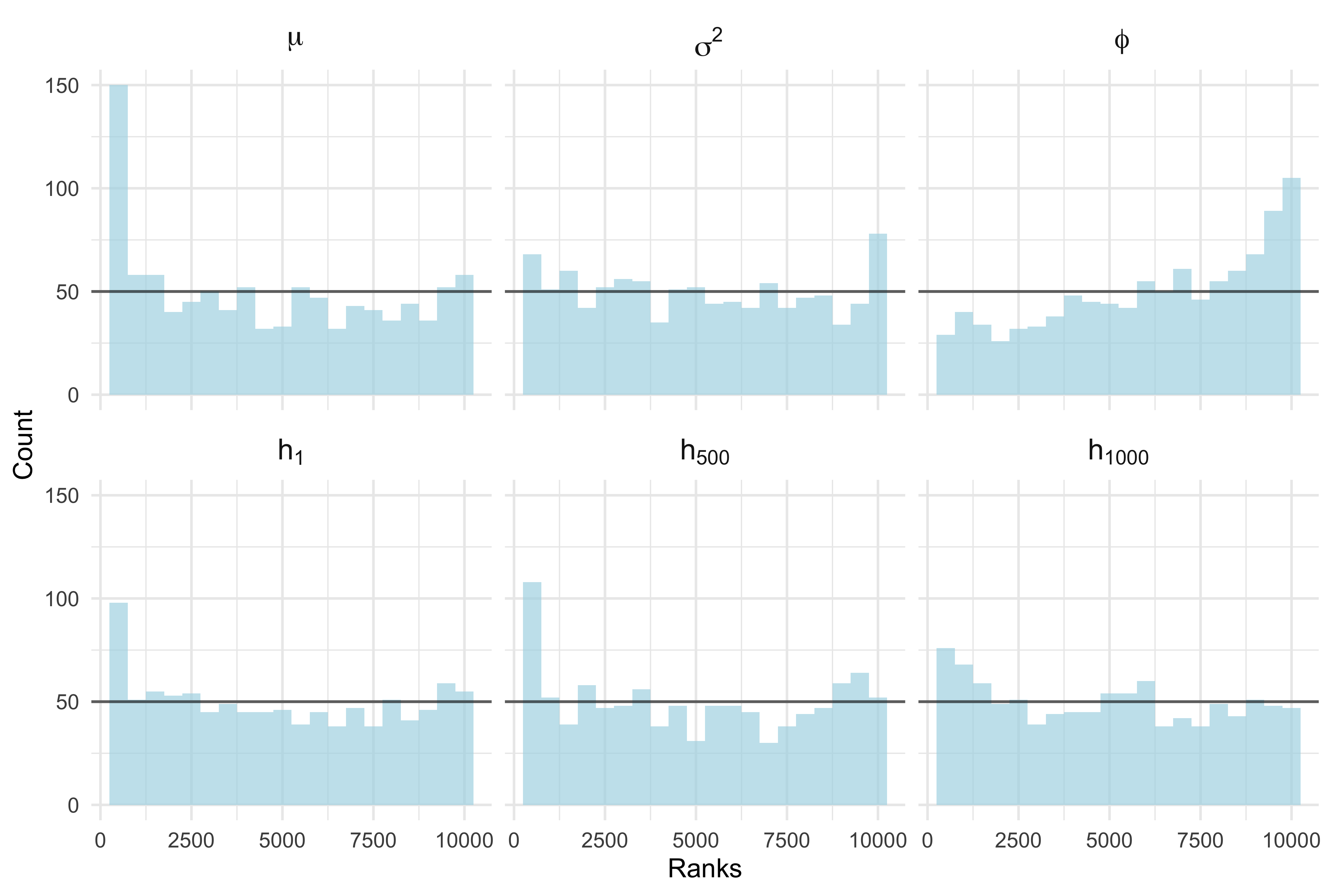}
        \caption{\textbf{1000 SBC iterations for centered Gaussian off-set mixture model}. $\mu$ has an inflated frequency on the left side suggesting the algorithm is over estimating the true parameter on average. The converse can be said about $\phi$ which possesses inflated frequency on the right hand side which implies underestimation of the true parameter.}
        \label{fig:cpksc1k}
    \end{figure}

    \begin{figure}[H]
        \centering
        \includegraphics[scale=0.09]{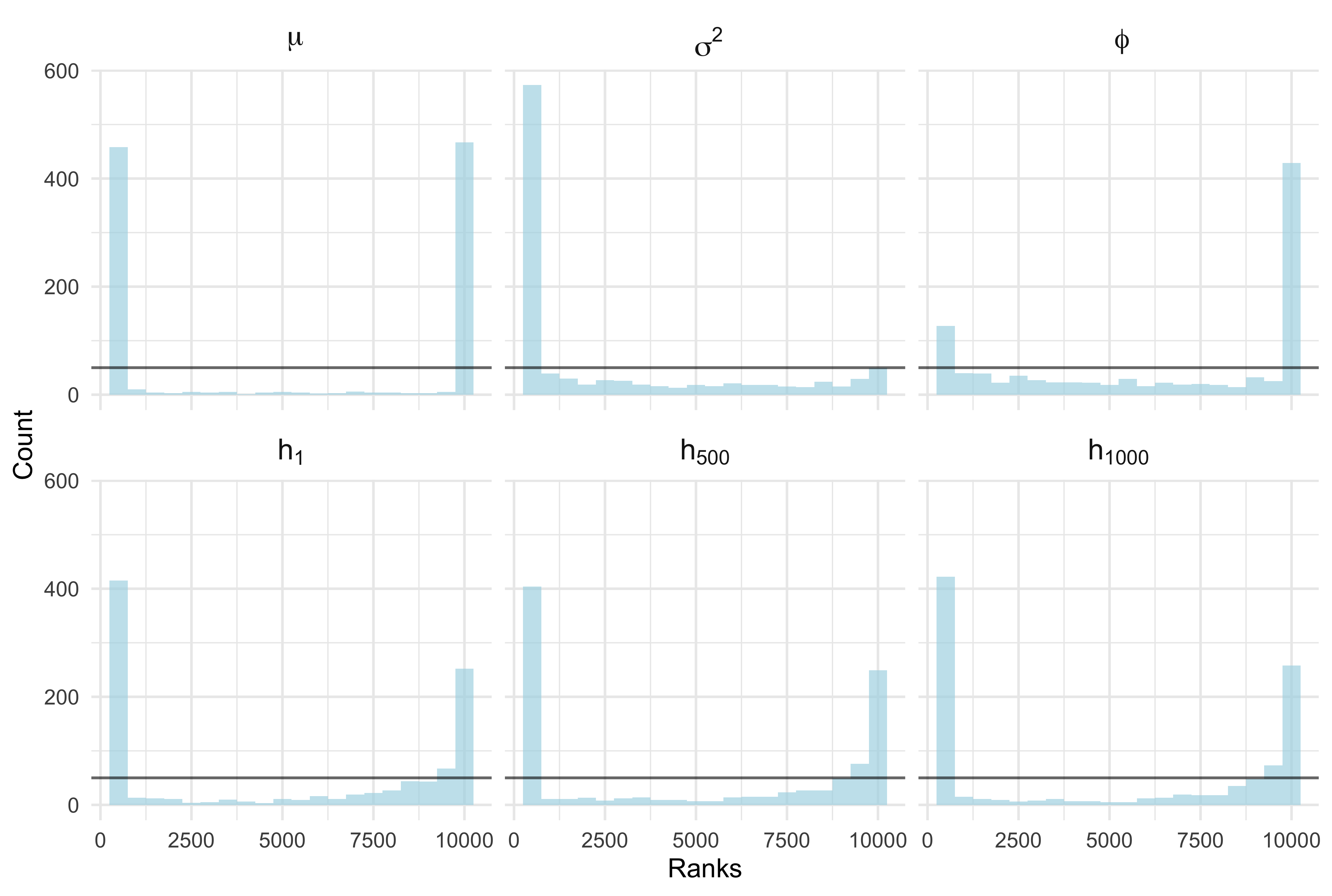}
        \caption{\textbf{1000 SBC iterations for non centered in location Gaussian off-set mixture model}. The rank statistics for all selected parameters are non uniform in shape. The KSC bespoke MCMC struggles to return the correct posteriors for this parameterisation of the model.}
        \label{fig:ncpksc1k}
    \end{figure}
    
\section{Appendix E: Importance weighted rank statistics for reparameterised model}

\begin{figure}[H]
        \centering
        \includegraphics[scale=0.09]{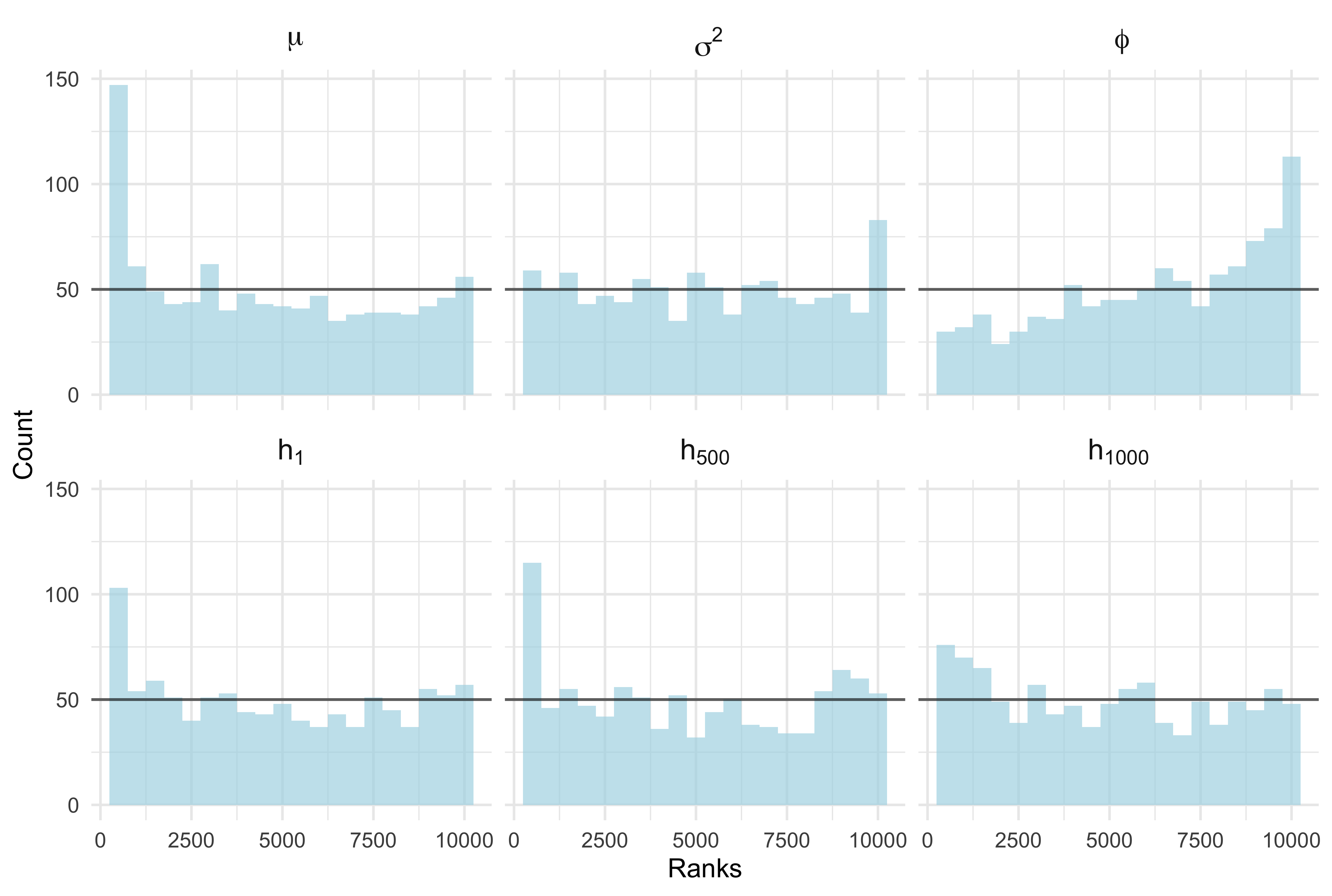}
        \caption{\textbf{1000 SBC iterations for re-weighted rank statistics from the off-set mixture model}.}
        \label{fig:reweight1k}
    \end{figure}

\begin{figure}[H]
    \centering
    \includegraphics[scale=0.09]{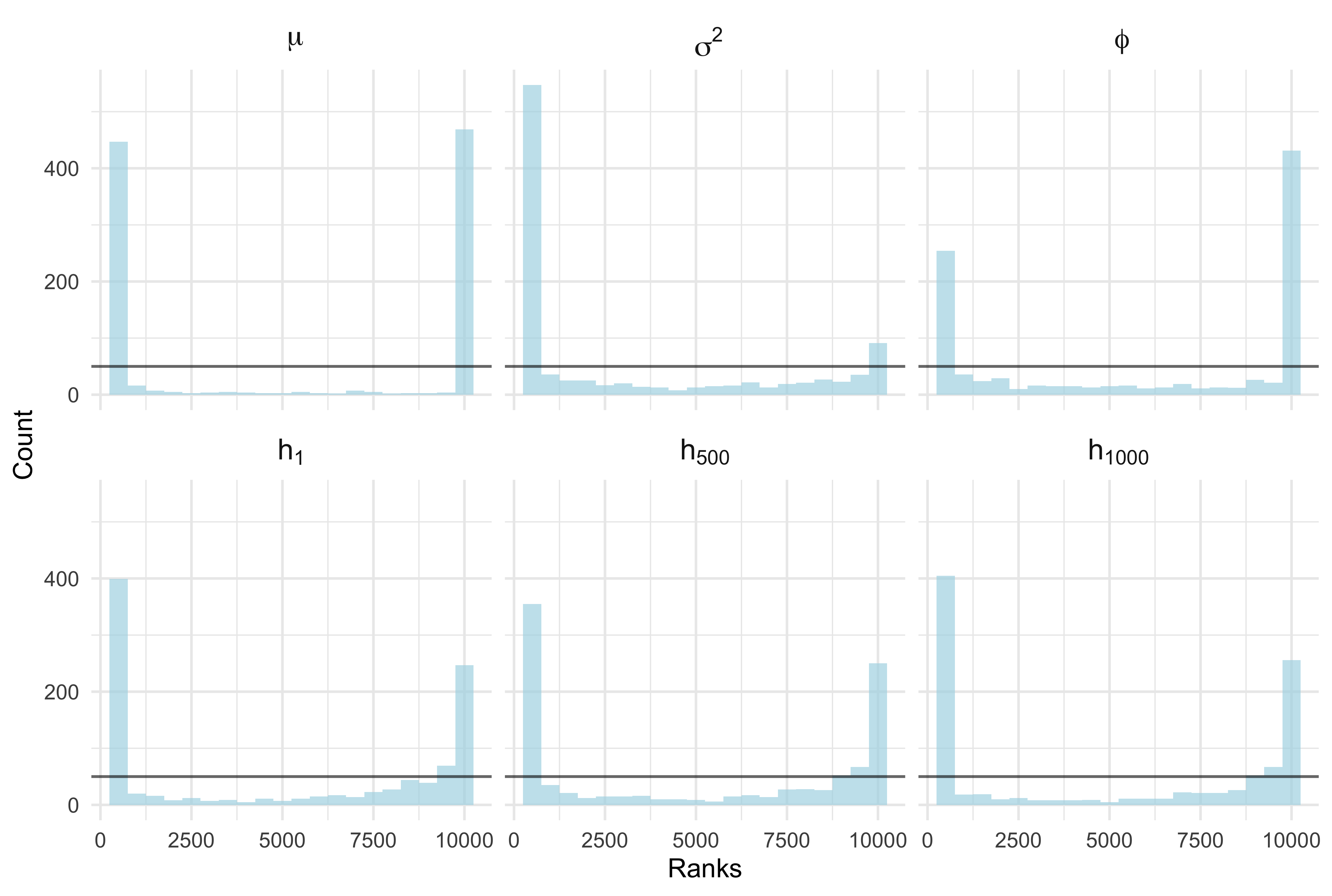}
    \caption{\textbf{1000 SBC iterations for re-weighted rank statistics from the reparameterised off-set mixture model}.}
    \label{fig:ncpreweight1k}
\end{figure}

\begin{figure}[H]
    \centering
    \includegraphics[scale=0.09]{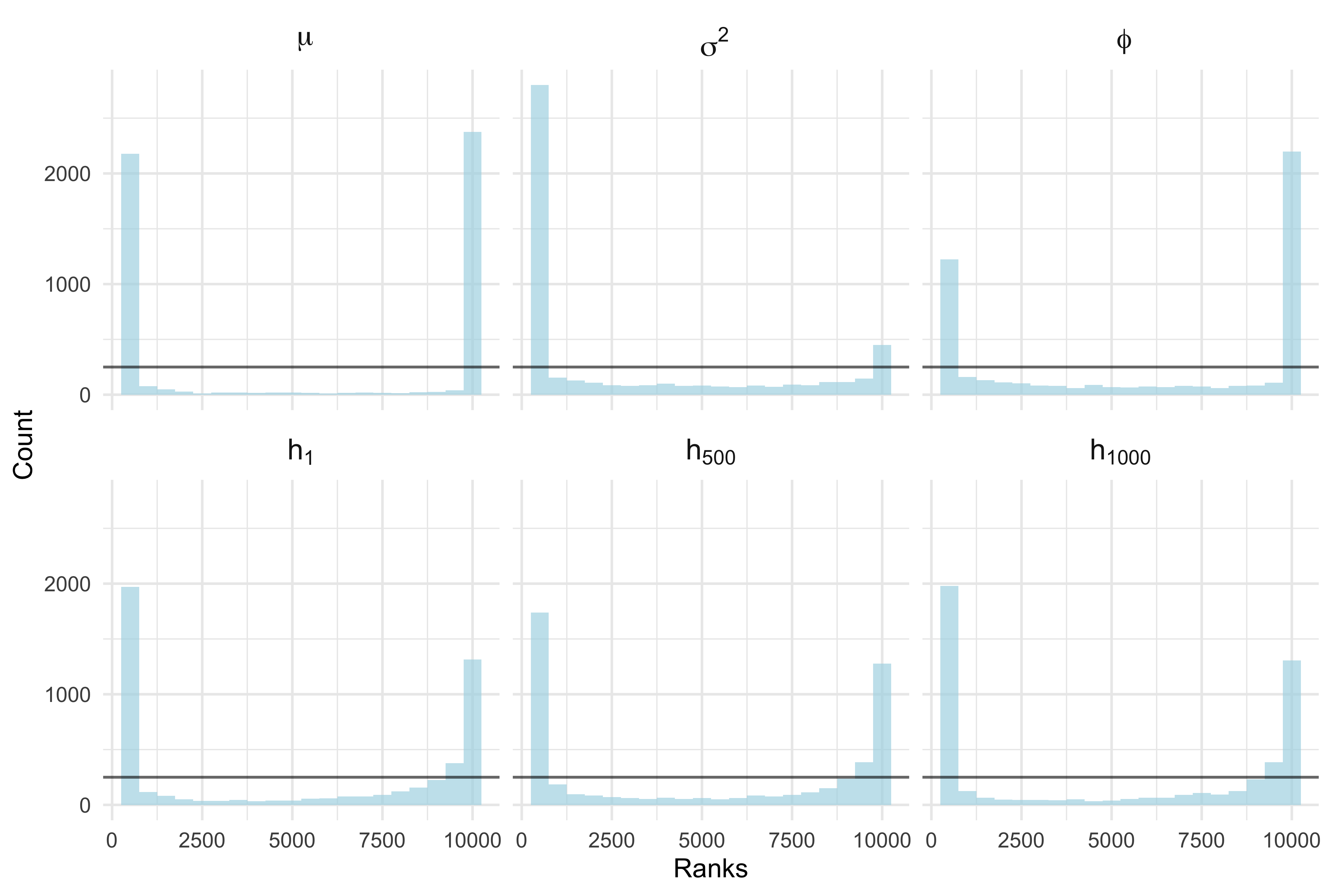}
    \caption{\textbf{5000 SBC iterations for re-weighted rank statistics from the reparameterised off-set mixture model.}}
    \label{fig:ncpreweight5k}
\end{figure}

\end{document}